# Combining rock typing and NMR logging to estimate permeability in complex carbonate formations

Nihal Malavalli Lokesh[1], Biswajit Thander[2], Esmail Ansari[3], Ali Eghbali[4], and Behzad Ghanbarian[1,5,6*]

[1] Division of Data Science, College of Science, University of Texas at Arlington, Arlington TX 76019, United States

[2] Department of Applied Geology, Dibrugarh University, Dibrugarh-786004, Assam, India

[3] Kansas Geological Survey, University of Kansas, Kansas, USA

[4] Baker Hughes, Houston, Texas, USA

[5] iResearrchE[3] Lab., Department of Earth and Environmental Sciences, University of Texas at Arlington, Arlington TX 76019, United States

[6] Department of Civil Engineering, University of Texas at Arlington, Arlington TX 76019, United States

[*] Corresponding author's Email address: ghanbarianb@uta.edu

**Abstract**

Estimating intrinsic permeability ($k$) from nuclear magnetic resonance (NMR) logs is difficult in heterogeneous carbonates because a single calibration may not represent distinct pore systems. We evaluated whether flow-zone-indicator (FZI)-based rock typing improves empirical calibration of the modified Schlumberger-Doll Research (SDR) model in the Arbuckle Group, Kansas. The analysis used 470 depth-matched core-permeability and NMR measurements from the Wellington and Cutter wells. FZI calculated from the measured $k$ and NMR-derived porosity was grouped using K-means clustering approach. Six clusters were identified, but one singleton cluster was excluded, leaving five rock types for SDR calibration. Relative to the global model (without clustering), rock-type-specific calibration reduced testing root mean square log-transformed error (RMSLE) from 1.215 to 0.240–0.427 and increased coefficient of determination (CD) from −0.106 to 0.578–0.924. The fitted $T_{2lm}$ exponents ranged from 0.053 to 0.521, indicating that $T_{2lm}$ provided limited additional information in some rock types and that the SDR parameters should be treated as empirical calibration quantities. Because measured $k$ was used to calculate the FZI before the within-cluster splits, these results represent conditional performance for samples with known FZI membership, not blind $k$ prediction. Field implementation will require a permeability-independent method for assigning rock types.

**Keywords:** Arbuckle formation, Nuclear magnetic resonance, Permeability, Porosity, Rock typing

## 1. Introduction

Formation characterization is essential for understanding spatial heterogeneity controlling groundwater flow and storage, which in turn influences subsurface water resources management. Therefore, various methods have been proposed for subsurface characterization, such as core sampling (Sudicky, 1986; Alexander et al., 2011), tracers (Geyer et al., 2007; Doro et al., 2015; Sarris et al., 2018), electrical resistivity tomography (Gomez-Nicolas et al., 2018; Robinson et al., 2016), ground penetrating radar (Bowling et al., 2005; Parker et al., 2022) and hydraulic tomography (Bohling and Butler, 2010; Berg and Illman, 2015; Sun et al., 2024). Among them, nuclear magnetic resonance (NMR) logging is a promising technique that provides direct measurements of liquid porosity and fluid content (Coates et al., 1999) and indirect estimation of pore size distribution (James and Ehrlich, 1999) and permeability (Parsekian et al., 2015). NMR applications have been widely and routinely used in petroleum engineering (Kenyon et al., 1988; Coates et al., 1999) and growing to aquifer characterization in subsurface hydrology (Knight et al., 2016; Ren et al., 2019; Kendrick et al., 2021; Phillips et al., 2024). In principles, NMR operates by detecting the response of hydrogen nuclei in water molecules subjected to an external magnetic field and their relaxation times (e.g., $T_2$). Depending on the external magnetic field direction, NMR measures longitudinal ($T_1$) or transverse ($T_2$) relaxation time distribution (Walsh et al., 2013).

Coates et al. (1999) investigated the link between permeability, $k$, and NMR $T_2$ relaxation time. They assumed that NMR $T_2$ distribution represents irreducible and free fluid separated by a cutoff $T_2$ value and proposed the following empirical relationship (known as Timur-Coates model in the literature):

$$k = \left(\frac{\phi}{C_{TC}}\right)^4 \left(\frac{FFI}{BVI}\right)^2 \quad (1)$$

where $\phi$ is the porosity, FFI is the free fluid index, BVI is the bulk volume of irreducible water, and $C_{TC}$ is a numerical prefactor (Timur-Coates constant) whose value depends on processes that created a formation and, thus, varies from one formation to another (Coates et al., 1999). For instance, Coates et al. (1999) reported $C_{TC}$ = 6 and 10 for two formations (see pages 124 and 150 of Coates et al., 1999). Eq. (1) was later modified as $k = \left(\frac{\phi}{C_{mTC}}\right)^m \left(\frac{FFI}{BVI}\right)^n$ in which exponents $m$ and $n$ should be calibrated along the numerical prefactor $C_{mTC}$ using experimental data (Thibidoux et al., 2009).

The parameters FFI, BVI and $\phi$ in Eq. (1) can be directly determined from NMR and $T_2$ distribution measurements. For instance, the values of BVI and FFI can be calculated by integrating over the $T_2$ distribution from its smallest value to some cutoff value (known as $T_{2cutoff}$) and from $T_{2cutoff}$ to the maximum $T_2$ value, respectively. $T_{2cutoff}$ is typically around 90 milliseconds for carbonates (Chi and Heidari, 2016) but a wider range from 18.3 to 1072 milliseconds has been reported in the literature (Chen et al., 2008). The constant coefficient $C_{TC}$, however, must be calibrated using experimental data, a challenge in estimating permeability from NMR data, particularly in heterogeneous aquifers composed of layers of different compositional environments. Unlike a universal constant, the $C_{TC}$ value varies significantly with different types of reservoirs, such as clean formations have different constant value than that of the formation having high clay contents (Tiab & Donaldson, 2011).

In addition to the Coates et al. (1999) model, Schlumberger Doll Research proposed another model (hereafter the SDR model) widely applied to estimate permeability from NMR $T_2$ distribution. The SDR model is (Kenyon et al., 1988)

$$k = C_{SDR}\phi^4 T_{2lm}^2 \quad (2)$$

where $T_{2lm}$ is the logarithmic mean of the NMR transverse-relaxation-time distribution, and $C_{SDR}$ is a numerical prefactor that depends on the formation type and lithology (Kenyon et al., 1988). The original SDR relationship, Eq. (2), was developed and evaluated primarily using water-saturated sandstones (Kenyon et al., 1988). Its application to heterogeneous carbonate formations is less straightforward. Carbonate pore systems commonly contain combinations of interparticle pores, microporosity, molds, vugs, fractures, and dissolution features. Moreover, variations in surface relaxivity and pore coupling can cause similar $T_2$ responses to correspond to substantially different hydraulic properties. Consequently, the classical SDR exponents should not be assumed to have universal physical meaning in carbonate formations, and even empirically optimized exponents may reflect both formation properties and statistical characteristics of the calibration dataset. More specifically, Hodgkins and Howard (1999) showed that applying standard SDR calibration parameters to low-resistivity Gulf of Mexico sands could produce substantial discrepancies between NMR-derived and independently estimated formation permeabilities, highlighting the need for formation-specific calibration. Therefore, the SDR model was later modified as follows (hereafter the modified SDR model)

$$k = C_{mSDR}\phi^{\alpha} T_{2lm}^{\beta} \quad (3)$$

where $C_{mSDR}$ is an empirical coefficient, and $\alpha$ and $\beta$ are the fitted porosity and $T_{2lm}$ exponents, respectively. These parameters require formation-specific calibration using measured permeability and NMR data (Parsekian et al., 2015; Knight et al., 2016; Ren et al., 2019). In heterogeneous carbonates, the fitted parameters should primarily be regarded as empirical calibration parameters unless their physical interpretations can be independently supported.

Although the SDR model and its modified forms have been calibrated using experimental and field data (Dlubac et al., 2013; Parsekian et al., 2015; Knight et al., 2016), limited attention has been given to whether calibration within individual hydraulic rock types improves their performance in heterogeneous carbonate formations. Accordingly, this study does not assume that the SDR relationship is universally applicable to carbonate rocks or that its fitted parameters necessarily possess unique physical meanings. Instead, the objectives are to: (1) identify rock types with distinct hydraulic characteristics, (2) calibrate the modified SDR relationship separately within each identified rock type, and (3) determine whether rock-type-specific calibration improves permeability estimation relative to a single global SDR calibration.

## 2. Materials and Methods

### 2.1. Study area

The study area is located in the state of Kansas, United States, with NMR well log data and core samples collected from two wells (i.e., Wellington and Cutter) in the Arbuckle formation, as shown in Fig. 1. The Arbuckle Group in the craton of North America consists mostly of carbonate with a few sandstone lenses. The Arbuckle limestone was identified based on the exposed rock strata in the Arbuckle Mountains of Precambrian to Simpson in southeastern Oklahoma of Late Cambrian and Early Ordovician age. It is evident from a drilled well near Caney, Kansas, that pre-Mississippian strata are present in Kansas (Schrader, 1908). Edson (1929) correlated Arbuckle limestone of in Kansas of Upper Cambrian and Ordovician in age with the surrounding states and found that Kansas Arbuckle limestone sandy lenses are different than other pre-Mississippian dolomites in Kansas.

Along the Wellington KGS 1-32 well (hereafter Wellington), located in the Sumner County, 405 core samples were collected from the depth of 4166.30 ft (1269.9 m) to 5162.70 ft (1573.6 m). Along the Cutter KGS-1 well (hereafter Cutter), located in the Stevens County, 289 core samples were collected between the depth of 5238.3 ft (1596.6 m) and 7587.7 ft (2312.7 m).

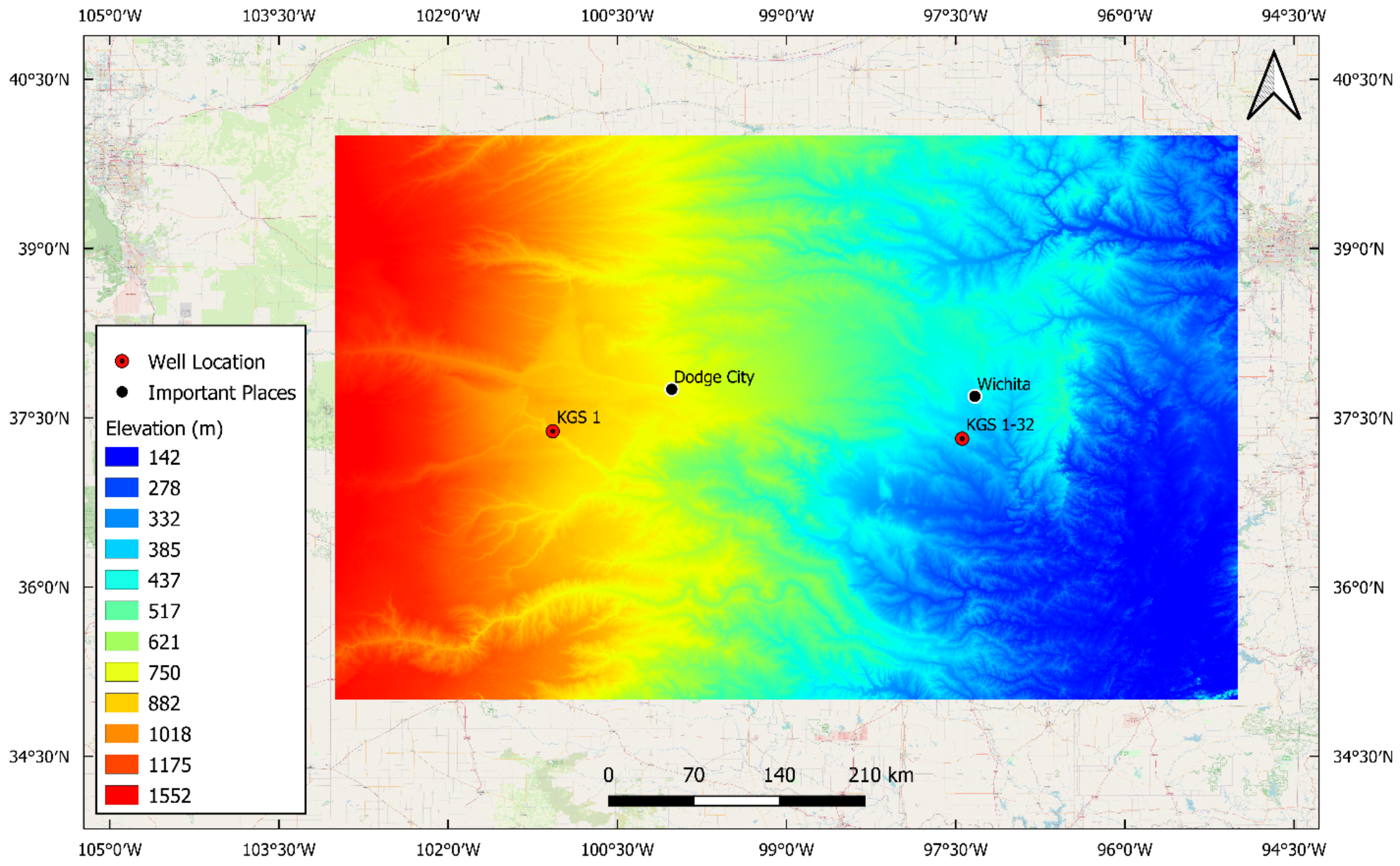

Figure 1. The locations of the Cutter and Wellington wells in the state of Kansas, United States.

Core samples collected from the Wellington and Cutter wells in the Arbuckle formation were analyzed for routine petrophysical properties, including porosity and permeability. Nuclear magnetic resonance (NMR) logging was conducted using modern wireline logging tools as part of an extensive petrophysical characterization program (Holubnyak et al., 2017). We cleaned and quality-controlled the database by matching the depths of the core measurements with the corresponding depths of the NMR $T_2$ distributions, resulting in a final dataset consisting of 470 samples.

### 2.2. Rock typing

Rock typing is a clustering method from petroleum engineering that groups rocks based on similarities in their hydraulic properties (Ghanbarian et al., 2019). Various rock-typing approaches have been proposed. A widely-used method, developed by Amaefule et al. (1993), identifies hydraulic flow units from porosity and permeability measurements using the following modified Kozeny–Carman relationship:

$$k = \frac{\phi^3}{(1-\phi)^2}\left[\frac{1}{F_s \tau^2 S_{gv}^2}\right] \quad (4)$$

where $F_s$ is the shape factor, $\tau$ is the tortuosity, and $S$ is the specific surface area (surface area per unit volume of rock).

Amaefule et al. (1993) rearranged Eq. (4) and defined the reservoir quality index (RQI), normalized porosity ($\phi_z$), also known as void ratio, and flow zone indicator (FZI) as follows

$$RQI = 0.0314\sqrt{\frac{k}{\phi}} \quad (5)$$

$$\phi_z = \frac{\phi}{1-\phi} \quad (6)$$

$$FZI = \frac{RQI}{\phi_z} \quad (7)$$

Each FZI represents a distinct flow unit in which pores have similar pore throat characteristics. In addition, a single flow unit can discontinuously occur anywhere across the depth profile of the formation.

In the Amaefule et al. (1993) approach, samples that align along a straight line with a unit slope on a log-log plot of RQI versus $\phi_z$ are interpreted to belong to the same hydraulic flow unit (HFU) or rock type. Each HFU is characterized by a distinct FZI, defined as the intercept of the unit-slope line at $\phi_z = 1$.

For each sample, the FZI was calculated using the measured permeability and NMR-derived porosity according to Eq. (7). The K-means clustering approach was then applied to log-transformed FZI values to identify samples belonging to distinct hydraulic flow units. Thus, rock typing was performed before dividing the samples within each cluster into calibration and testing subsets. Because measured permeability is required to calculate FZI, the rock-type assignments used in this study were known a priori and were not independently predicted from NMR measurements. Because the optimal number of clusters was not known a priori, we used the Elbow method (James et al., 2013) to determine the appropriate number of clusters. Specifically, the within-cluster sum of squares was calculated for cluster numbers ranging from 1 to 15, and the optimal cluster number was selected based on the location of the elbow in the resulting curve.

### 2.3. Calibrating the modified SDR model parameters

In carbonate reservoirs, complexities in the pore space are because of primary and post depositional secondary porosities, such as interparticle, inter-crystalline, microporosity and vugs (Moshier, 1989; Moslemipour and Sadeghnejad, 2021). In addition to that, carbonate reservoirs are spatially very heterogeneous (Lucia, 1999). Such spatial heterogeneities typically result in weak correlation between porosity and permeability and, therefore, calibrating the modified SDR model, Eq. (3), within each rock type becomes necessary.

To calibrate the modified SDR model parameters, i.e., $C_{MSDR}$, $\alpha$ and $\beta$, we linearized Eq. (3) using the logarithmic transformation as follows

$$log\ (k)\ = \log(C_{MSDR}) + \alpha log\ (\phi_{NMR})\ + \beta log(T_{2lm}) \quad (8)$$

Equation (8) allowed us to estimate $C_{MSDR}$, $\alpha$ and $\beta$ simultaneously by directly fitting it to the data using a multiple linear regression, given that $k$, $\phi_{NMR}$ and $T_{2lm}$ values were experimentally measured and available. Following rock typing, the samples within each sufficiently populated cluster were randomly divided once into fixed calibration and testing subsets containing approximately 80% and 20% of the samples, respectively. The testing subset was held fixed throughout the analysis. For each of 10,000 bootstrap iterations, following Parsekian et al. (2015), a sample equal in size to the calibration subset was drawn with replacement from that subset, and the modified SDR parameters were estimated from the resulting bootstrap sample. Each fitted model was evaluated on both its corresponding bootstrap sample and the fixed testing subset. The statistical error parameters were calculated separately for every iteration, and their averages across the 10,000 iterations were used to summarize model performance. The resulting training statistics represent apparent performance on the bootstrap samples, whereas the testing statistics quantify the transferability of the bootstrap-calibrated relationships to the fixed held-out subset within each predefined FZI rock type.

To provide a representative visualization, we selected the realization whose training RMSLE was closest to the mean training RMSLE across the 10,000 realizations. The fitted parameters and data partition from that realization were then used to present both the training and corresponding testing results. Testing performance was not considered when selecting the representative realization.

### 2.4. Model performance

To evaluate the performance of the modified SDR model, we calculated the root mean square log-transformed error (RMSLE) and coefficient of determination (CD) defined as

$$RMSLE = \sqrt{\frac{\sum_{i=1}^{N}[log\,(k_{est}^{i}) - log\,(k_{meas}^{i})]^2}{N}} \tag{9}$$

$$CD = 1 - \frac{SS_{res}}{SS_{tot}} \tag{10}$$

where $N$ is the number of samples, $k_{est}$ and $k_{meas}$ are the estimated and measured permeability values, respectively, $SS_{res}$ is the residual sum of squares, and $SS_{tot}$ is the total sum of squares. We should note that the coefficient of determination was calculated in logarithmic permeability space because permeability spans several orders of magnitude. Thus, $SS_{\mathrm{res}}$ and $SS_{\mathrm{tot}}$ were calculated using $\log(k)$.

## 3. Results and Discussion

In this section, we first present the results of optimizing the modified SDR model parameters using the entire dataset without applying rock typing. We then present the results of the rock typing analysis together with the optimized modified SDR model parameters obtained for each hydraulic flow unit cluster separately.

### 3.1. Optimizing the modified SDR model parameters without rock typing

Figure 2 shows the distribution of the optimized model parameters – i.e., $log\ (C_{mSDR})$ , $\alpha$, and $\beta$ – obtained using the bootstrapping method with 10,000 realizations. As can be seen, all histograms (Figs. 2a-c) show bell-shaped patterns with the average values of $C_{mSDR}$ = 0.001±0.000 mD ms$^{-\beta}$, $\alpha$ = 0.924±0.191, and $\beta$ = 1.469±0.156.

Figures 2d-e display the estimated permeability against the measured value for the training and testing datasets for a representative realization. In both plots, the dashed red line represents the 1:1 line. Across the 10,000 bootstrap iterations, the mean RMSLE values were 1.058 and 1.215 and the mean CD values were 0.311 and −0.106 for the bootstrap training samples and fixed testing subset, respectively. Figures 2d and 2e show one representative bootstrap realization selected using the criterion described in Section 2.3; consequently, its realization-specific statistics may differ slightly from these mean values. Systematic errors are clear in both training and testing datasets; permeability is mainly overestimated at lower $k$ values (< 2 mD), while underestimated at higher $k$ values (> 2 mD). Similar overestimations by the SDR model at low $k$ values (< 1 mD) were reported by Cheng and Heidari (2015). Recently, Ghanbarian et al. (2026) highlighted several important sources of uncertainty that can prevent the modified SDR model parameters from being uniquely or perfectly optimized across heterogeneous rock samples. They argued that the modified SDR parameters are strongly influenced by lithology-dependent properties, particularly variability in surface relaxivity, pore geometry, pore connectivity, and the Archie cementation exponent (Archie, 1942). Because surface relaxivity depends on mineralogy, surface chemistry, wettability, and the presence of paramagnetic ions, its value may vary substantially among samples and lithologies, making a single global calibration

unreliable. Additional uncertainty arises from the selection of the characteristic NMR relaxation time since the NMR $T_2$ distribution reflects a complex mixture of pore-body and pore-throat signatures (Kewen and Ning, 2008) rather than a unique throat size distribution. Ghanbarian et al. (2026) further emphasized that the porosity exponent $\alpha$ may vary among rock types because it is theoretically linked to the cementation exponent, which itself depends on pore structure and connectivity. Consequently, calibrating a single set of modified SDR model parameters for heterogeneous datasets can introduce systematic bias and significant prediction scatter.

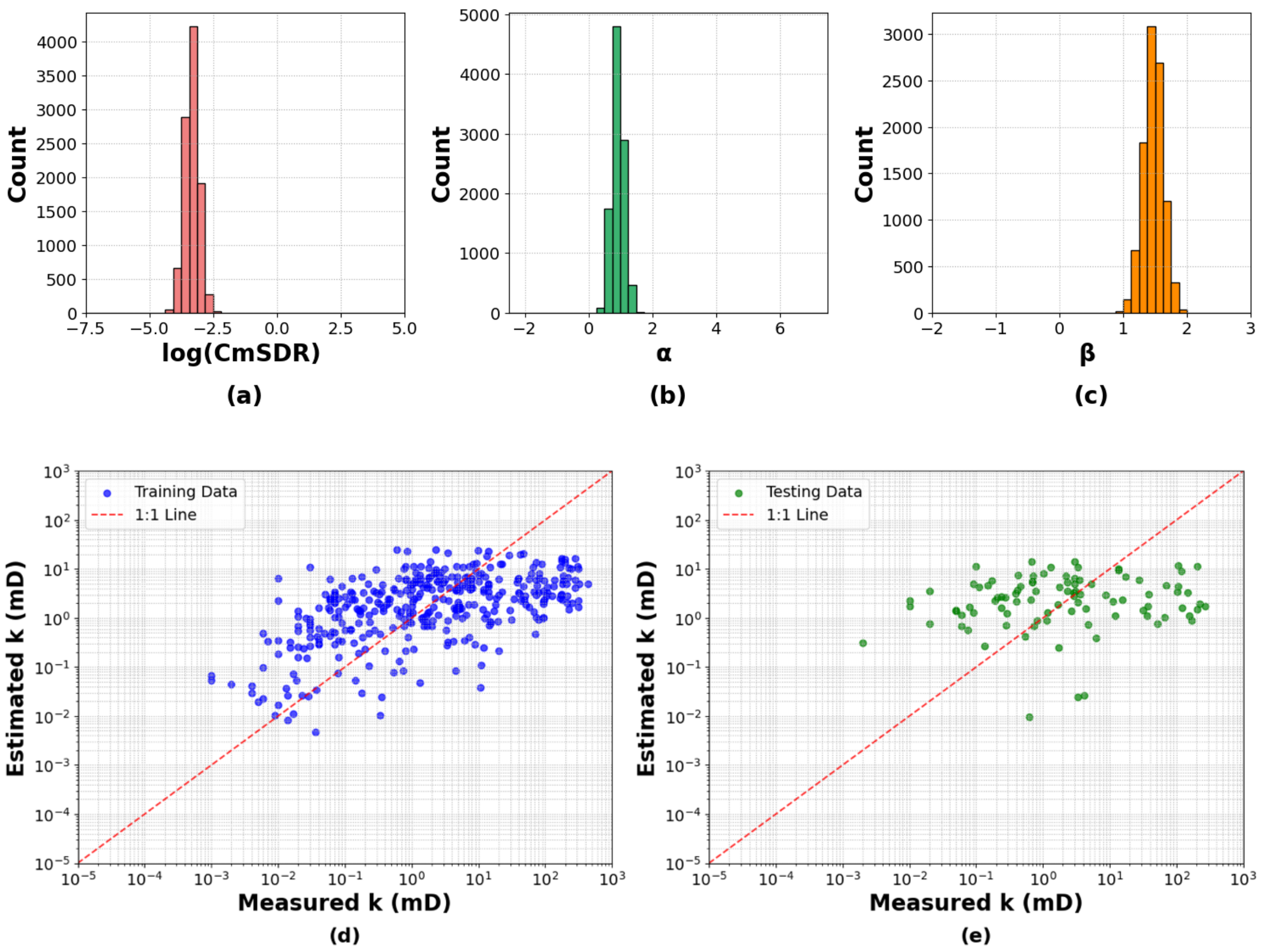

Figure 2. (a–c) Distributions of the modified SDR parameters obtained from 10,000 bootstrap samples drawn from the fixed calibration subset. (d, e) Estimated versus measured

permeability for the bootstrap training sample and fixed testing subset, respectively, for the representative realization selected as described in Section 2.3. The dashed red line represents 1:1 agreement.

### 3.2. Detected rock types

Figure 3(a) presents the trend between core permeability and core porosity measured on the 470 core samples on the semi-log scale. As can be seen, the data are weakly correlated and widely scattered, with $k$ values spanning nearly five orders of magnitude (from approximately $10^{-2}$ mD to over $10^{2}$ mD) for a relatively narrow range of core porosity values, particularly in the range of 0.02 and 0.05 (2-5%). Similar weak correlations between porosity and permeability have been reported for carbonate samples (see e.g., Babadagli and Al-Salmi, 2004). In carbonate rocks, the pore system is strongly controlled by depositional fabric and subsequent diagenetic overprinting, which can create secondary porosity types such as vugs, molds and fractures that contribute significantly to total porosity but may or may not be connected to the flow network (Lucia, 1983). As highlighted by Lønøy (2006), the pore type classification in carbonates is critical because interparticle porosity, intraparticle porosity, and vuggy porosity each exhibit distinctly different permeability behavior at equivalent porosity values. Furthermore, in tight carbonate formations with low matrix porosity (as seen in the cluster between 0.02 and 0.05), the permeability is often controlled by micro-fractures and pore throat connectivity rather than total pore volume, making standard porosity-permeability transforms unreliable (Aguilera, 2004). The poor correlation shown in Fig. 3a further validates the need for a rock typing approach that honors the pore-scale heterogeneity of the reservoir rather than relying on empirical global transforms.

Figure 3(b) shows the cross-plot of the NMR-derived porosity ($\phi_{NMR}$) against the core-measured porosity, with a 1:1 reference line (shown in red) indicating disagreement between the two measurements. The NMR-derived porosity was generally greater than core-measured porosity, indicating a predominantly positive—not random—difference between the two measurements. Differences in measurement support provide one plausible explanation: the NMR log samples a substantially larger formation volume and may detect vugs, fractures, dissolution features, and spatially distributed microporosity that are not represented by individual core plugs. Nevertheless, a difference in support volume alone does not necessarily produce a consistently positive bias. Partial gas or hydrocarbon saturation also does not provide a straightforward explanation because fluids with hydrogen indices below that of water generally reduce, rather than increase, apparent NMR porosity. Other possible influences include depth-matching uncertainty, environmental and borehole corrections, differences in measurement conditions, and the treatment of short-$T_2$ signal components. Because the information required to quantify these effects is unavailable, the observed positive difference remains an unresolved source of uncertainty.

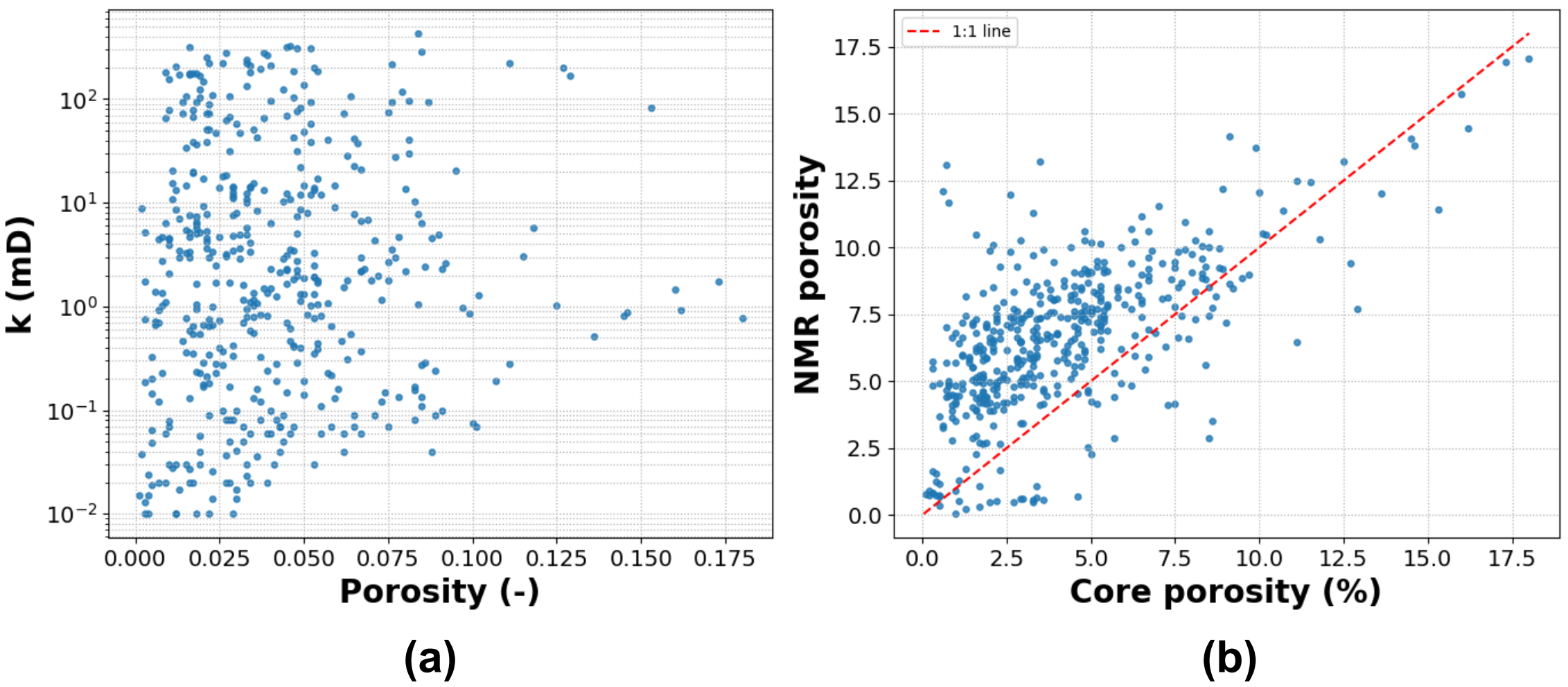

Figure 3. (a) Core permeability and (b) NMR porosity versus core porosity for 470 samples from the Arbuckle formation (Wellington and Cutter wells) used in this study. The red dashed line in (b) represents the 1:1 line.

The resulting clusters are shown on the log-log plot of reservoir quality index (RQI) versus normalized porosity ($\phi_z$) in Fig. 4b, where each cluster is represented by a different color together with its corresponding unit-slope regression line. Although six clusters were identified, the singleton sixth cluster was excluded; therefore, five rock types were modeled. Because the RQI and $\phi_z$ are related through the FZI according to Eq. (7), each parallel unit-slope line represents a constant FZI value and therefore a distinct hydraulic rock type. Clusters located toward the upper-left portion of the cross plot (Fig. 4b) correspond to larger FZI values, indicating better-connected pore systems and higher permeability for a given porosity, whereas clusters toward the lower-right represent tighter rock with poorer pore connectivity and lower flow capacity. More specifically, we found FZI = 0.411, 1.658, 5.125, 15.407, and 61.011, respectively, for Clusters 1 through 5.

As shown in Fig. 4b, Cluster 6 is defined by only a single data point at the low-$\phi_z$ end of the trend, with no other samples falling along its regression line. Because a single sample cannot support a statistically meaningful calibration of the modified SDR model or provide any basis for assessing prediction error, Cluster 6 was excluded from subsequent permeability modeling. Consequently, only Clusters 1 through 5 were carried forward for the modified SDR calibration and validation.

Figure 4b also demonstrates substantial variability in both RQI and $\phi_z$, spanning several orders of magnitude. The broad scatter indicates the presence of multiple rock types with significantly different pore-throat geometries and hydraulic characteristics within the Arbuckle formation. Such heterogeneity is typical of carbonate systems, where diagenetic processes including dissolution, cementation, dolomitization, and fracturing strongly modify pore structure independently of bulk porosity (Lucia, 1995). The absence of a single well-defined trend further suggests that porosity alone is insufficient to reliably predict permeability across the entire reservoir. Instead, the variability in the RQI-$\phi_z$ relationship highlights the importance of hydraulic rock typing to group samples with similar pore structure and flow behavior prior to permeability estimation (Corbett and Potter, 2004).

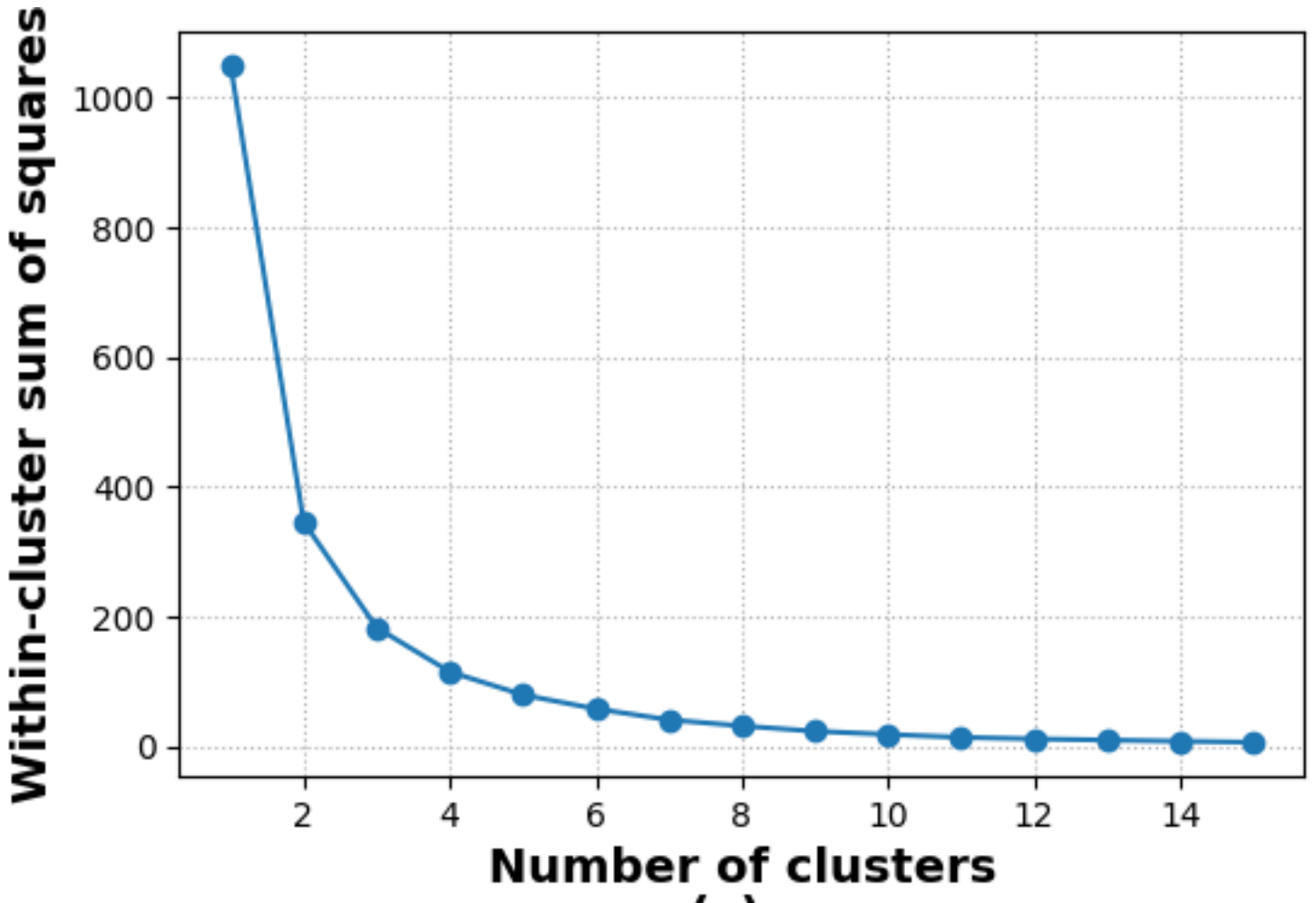

**(a)**

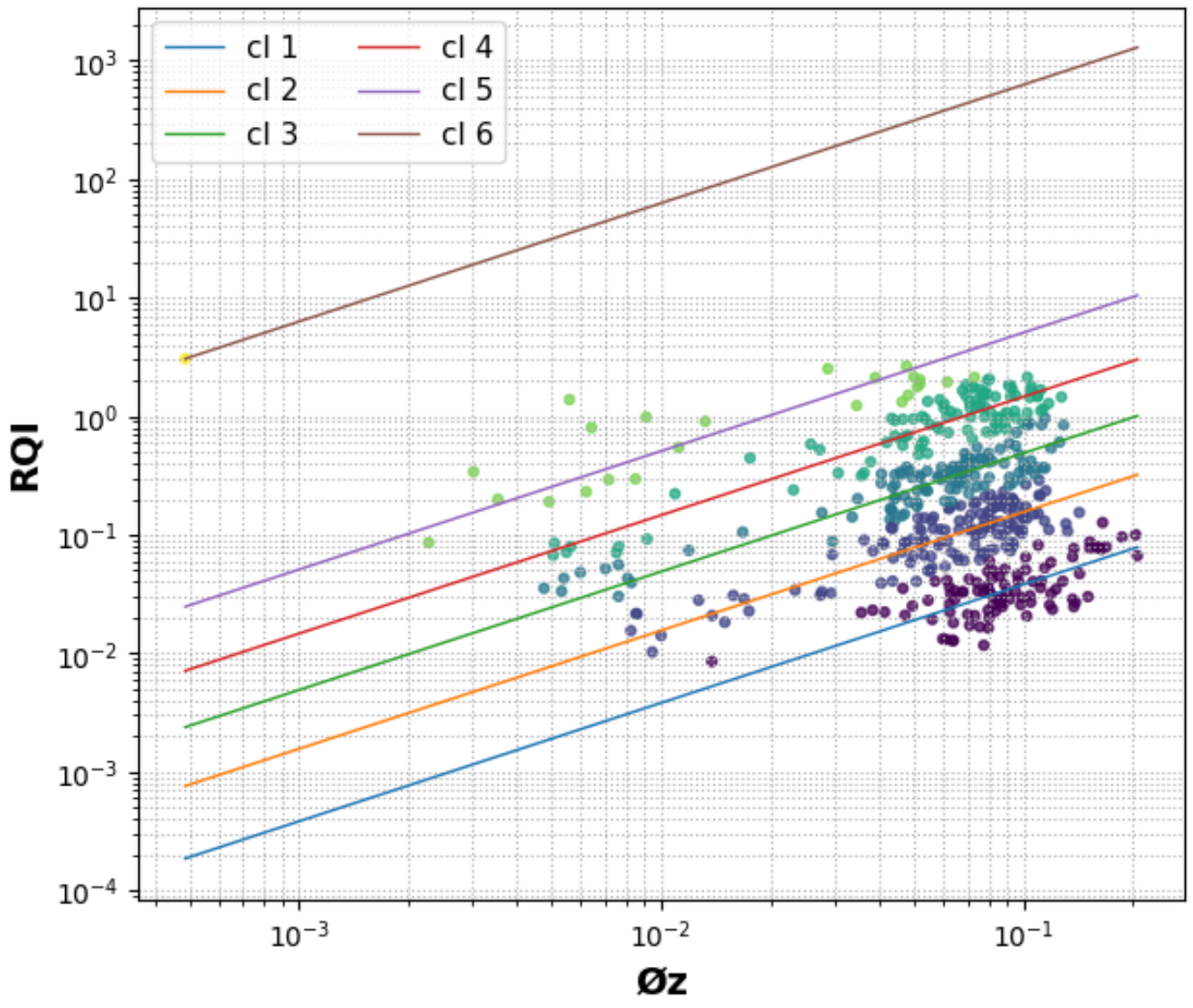

**(b)**

Figure 4. (a) Elbow plot of within-cluster sum of squares versus number of clusters used to determine the selected K = 6 for K-means clustering on log(FZI), and (b) the resulting six hydraulic flow unit clusters displayed on a log-log RQI versus $\phi_z$ cross plot with corresponding regression lines.

Figure 5 presents the relationship between core permeability and NMR porosity for the identified HFUs. The scatter points represent the laboratory-measured core data, while the solid curves correspond to the fit of the modified Kozeny-Carman model, Eq. (4), with the mean FZI value assigned to each cluster. The theoretical curves show good agreement with the measured data within their respective HFUs and generally pass through the center of the corresponding data clouds, indicating that the cluster-specific FZI values successfully capture the dominant flow characteristics of each rock type. Fig. 5 also demonstrates a strong separation among the HFUs, confirming that a single porosity-permeability relationship is insufficient to describe the heterogeneous carbonate reservoir. At a given NMR porosity, $k$ varies by several orders of magnitude depending on the assigned cluster, reflecting substantial differences in pore-throat geometry and hydraulic connectivity among the rock types. Nevertheless, within each individual cluster, permeability exhibits a clear and systematic increase with NMR porosity, consistent with established porous-media flow behavior. Overall, the results indicate that porosity alone is not a reliable predictor of permeability in heterogeneous carbonate formations. In contrast, incorporating pore-geometry effects through FZI-based rock typing effectively partitions the highly scattered global dataset into distinct and more predictable permeability trends, thereby improving reservoir characterization and permeability estimation.

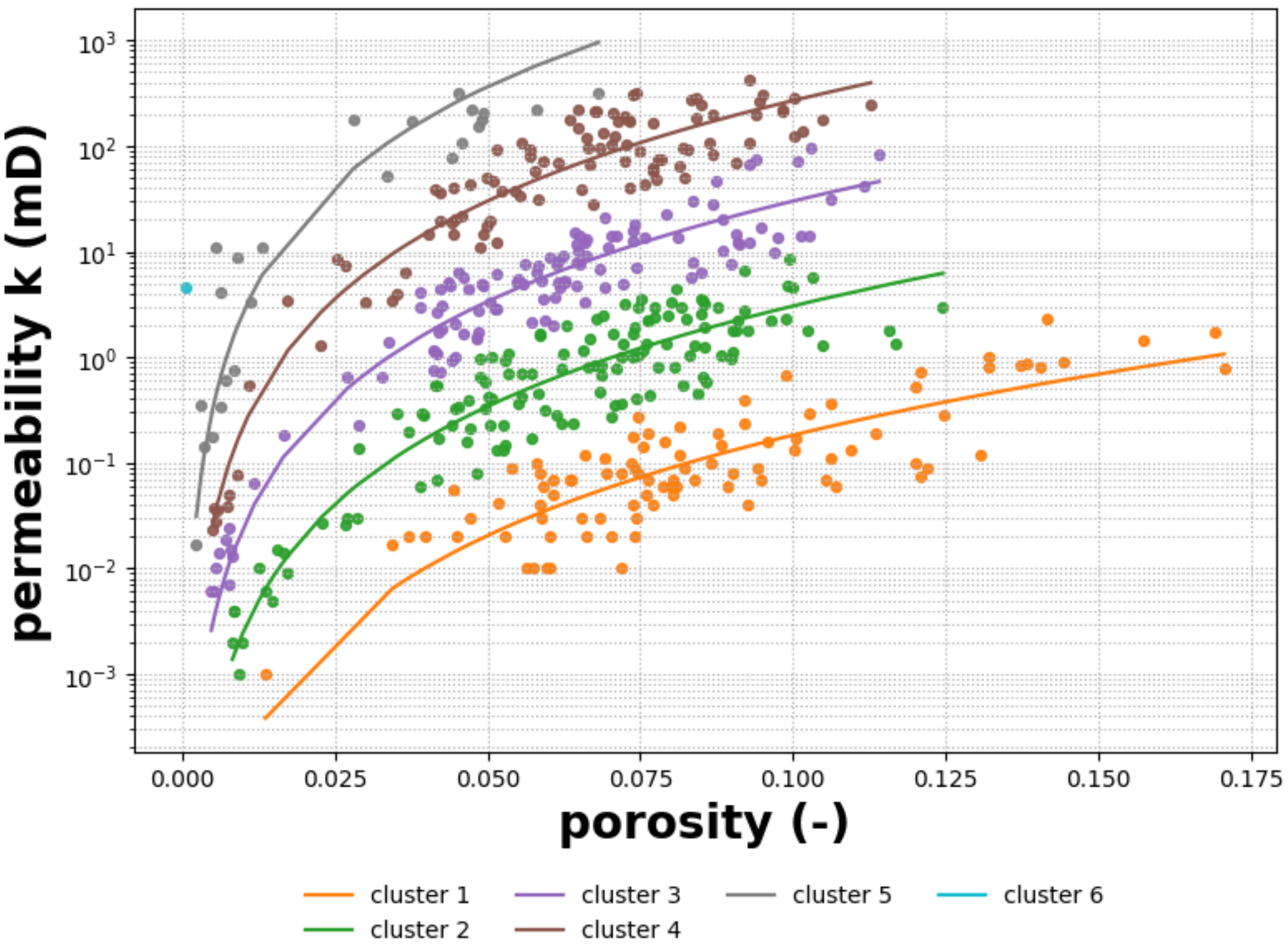

Figure 5. Core permeability versus NMR porosity and the corresponding fit of the modified Kozeny-Carman model, Eq. (4), for five hydraulic flow unit clusters illustrating the distinct permeability-porosity relationships captured within each cluster.

To evaluate the sensitivity of rock typing and SDR calibration to the porosity measurement, the workflow was also implemented using core-measured porosity in place of NMR-derived porosity. Those results are presented in the Supporting Information and discussed in Section 3.5. The two porosity inputs produced different FZI distributions and different selected numbers of clusters, demonstrating that hydraulic-flow-unit identification and the resulting SDR calibration are sensitive to the porosity dataset used. Therefore, rock types established from core measurements cannot necessarily be transferred directly to log-derived porosity without accounting for the systematic difference between the two measurements.

### 3.3. Optimizing the modified SDR model parameters with rock typing

The distributions of the optimized modified SDR model parameters for the five hydraulic rock types are presented in Figure 6, while the corresponding average values of $C_{mSDR}$, $\alpha$, $\beta$, and FZI are summarized in Table 1. Considerable variability in the fitted parameters was observed among the clusters, demonstrating the non-universality of the modified SDR relationship across the heterogeneous Arbuckle carbonates. The optimized $C_{mSDR}$ values ranged from 0.0001 to 2.2177 mD $\mathrm{ms}^{-\beta}$, whereas the porosity exponent, $\alpha$, ranged from 2.163 to 2.964. The fitted $T_{2lm}$ exponent, $\beta$, ranged from 0.053 to 0.521, substantially below the fixed value of 2 in the classical SDR model, Eq. (2). The classical exponents of 4 for porosity and 2 for $T_{2lm}$ were established from relationships involving core-measured permeability and porosity for water-saturated sandstone samples (Kenyon et al., 1988). In contrast, the present calibration combined core-measured permeability with NMR-derived porosity and was applied to heterogeneous carbonate rock types. Therefore, the departures from the classical exponents may reflect differences in lithology, pore structure, and porosity measurement, as well as correlations and uncertainty among the fitted parameters. No systematic relationship was observed between FZI and either $\alpha$ or $\beta$. Accordingly, the optimized parameters should be interpreted primarily as empirical, rock-type-specific calibration quantities rather than uniquely identifiable physical descriptors of the pore system.

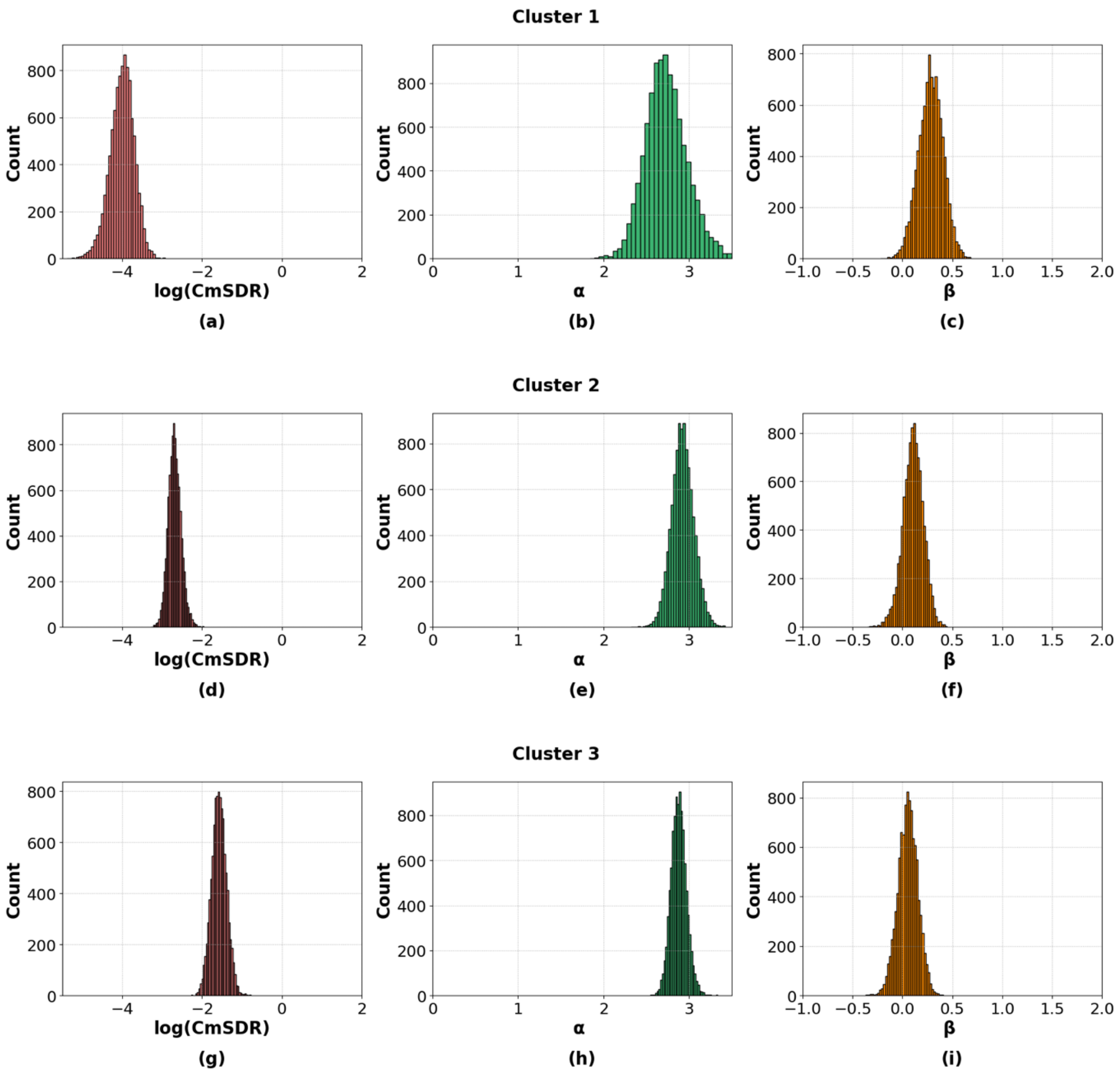

Figure 6. Distribution of the optimized SDR model parameters, including $log(C_{mSDR})$, $\alpha$, and $\beta$, showing the frequency and stability of the parameter values obtained during model calibration for each cluster and rock type.

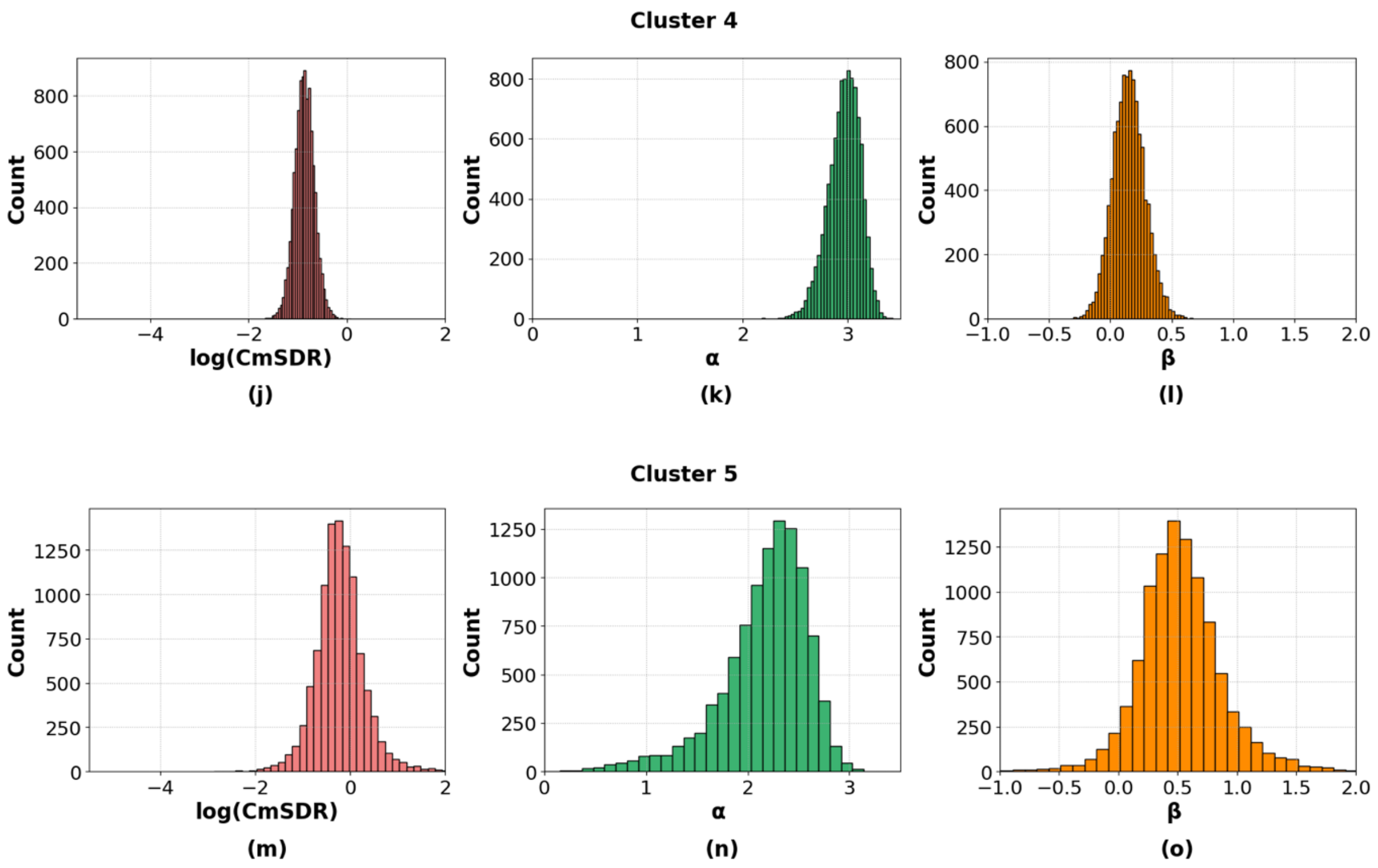

Figure 6. Continued.

Results presented in Table 1 indicate that the optimized $C_{mSDR}$ values generally increase with increasing FZI from Cluster 1 to 5. This trend suggests that the modified SDR prefactor is strongly influenced by hydraulic pore quality and pore-throat connectivity represented by the FZI. In contrast, no clear or systematic relationship was observed between the FZI and the optimized exponents $\alpha$ or $\beta$, indicating that these parameters are likely influenced by additional factors beyond hydraulic flow unit classification, including pore geometry complexity, surface relaxivity variability, cementation effects, and uncertainties associated with the characteristic $T_{2lm}$ relaxation time.

Table 1. The modified SDR model parameters optimized using carbonate rock samples from the Arbuckle formation, Kansas.

| Group | No. of samples | $C_{mSDR}$ (mD ms$^{-\beta}$) | $\alpha$ | $\beta$ | FZI |
|---|---|---|---|---|---|
| No Clustering | 470 | 0.0006 ± 0.000 | 0.924 ± 0.191 | 1.469 ± 0.156 | N/A |
| Cluster 1 | 95 | 0.0001 ± 0.000 | 2.734 ± 0.247 | 0.282 ± 0.123 | 0.411 ± 0.153 |
| Cluster 2 | 134 | 0.0022 ± 0.001 | 2.923 ± 0.130 | 0.105 ± 0.103 | 1.658 ± 0.562 |
| Cluster 3 | 116 | 0.0289 ± 0.014 | 2.877 ± 0.090 | 0.053 ± 0.100 | 5.125 ± 1.527 |
| Cluster 4 | 100 | 0.1506 ± 0.075 | 2.964 ± 0.152 | 0.149 ± 0.131 | 15.407 ± 4.896 |
| Cluster 5 | 24 | 2.2177 ± 22.689 | 2.163 ± 0.433 | 0.521 ± 0.354 | 61.011 ± 48.456 |
| Cluster 6 | 1 | - | - | - | - |

Clusters 2 and 3 yielded the smallest $\beta$ values, 0.105 and 0.053, respectively. Similar deviations from the classical SDR exponents have been previously reported in complex lithologies including carbonates, shales, and fractured rocks, where the assumptions underlying the original SDR formulation are not fully satisfied (Zhang et al., 2018; Mashhadi et al., 2025). The values of 0.105 and 0.053 are also substantially smaller than the fixed exponent of 2 in the classical SDR model, indicating that $T_{2lm}$ provided little additional explanatory contribution to permeability after accounting for porosity within these clusters. This result raises an important limitation concerning the applicability of the SDR functional form to these hydraulic rock types. In complex carbonate pore systems, $T_{2lm}$ may not adequately represent the pore throats and connected pathways controlling fluid flow because the measured relaxation distribution is also affected by pore-body size, surface relaxivity, pore coupling, and the coexistence of multiple pore types. Accordingly, the small fitted $\beta$ values should not be assigned a unique geological interpretation. They may reflect limited sensitivity of permeability to $T_{2lm}$, correlations among the fitted predictors, restricted within-cluster variability, parameter uncertainty, or a fundamental limitation of representing permeability through the SDR power-law form for these rock types.

Ghanbarian et al. (2026) theoretically proposed that, under specific assumptions applicable to Archie-type pore systems (Archie, 1942), $\alpha$ and $\beta$ should correspond to the cementation exponent and 2, respectively. The substantially smaller $\beta$ values obtained here demonstrate that those assumptions are not fully satisfied by the analyzed Arbuckle carbonates. Possible causes include variations in surface relaxivity, complex and multimodal pore geometry, incomplete correspondence between pore-body-sensitive NMR relaxation and flow-controlling pore throats, correlations among fitted parameters, and uncertainty in representing the entire $T_2$ distribution by a single characteristic value. Thus, the fitted exponents should not be interpreted as uniquely identifiable physical properties.

Figure 7 presents the comparison between laboratory-measured permeability and permeability estimated using the cluster-specific modified SDR models for the training and testing datasets. Overall, the results demonstrate that the rock-type-based calibration substantially improved $k$ estimation compared to the non-clustered approach (shown in Figs. 2d-e) by partitioning the heterogeneous carbonate reservoir into hydraulic flow units with more internally consistent petrophysical behavior.

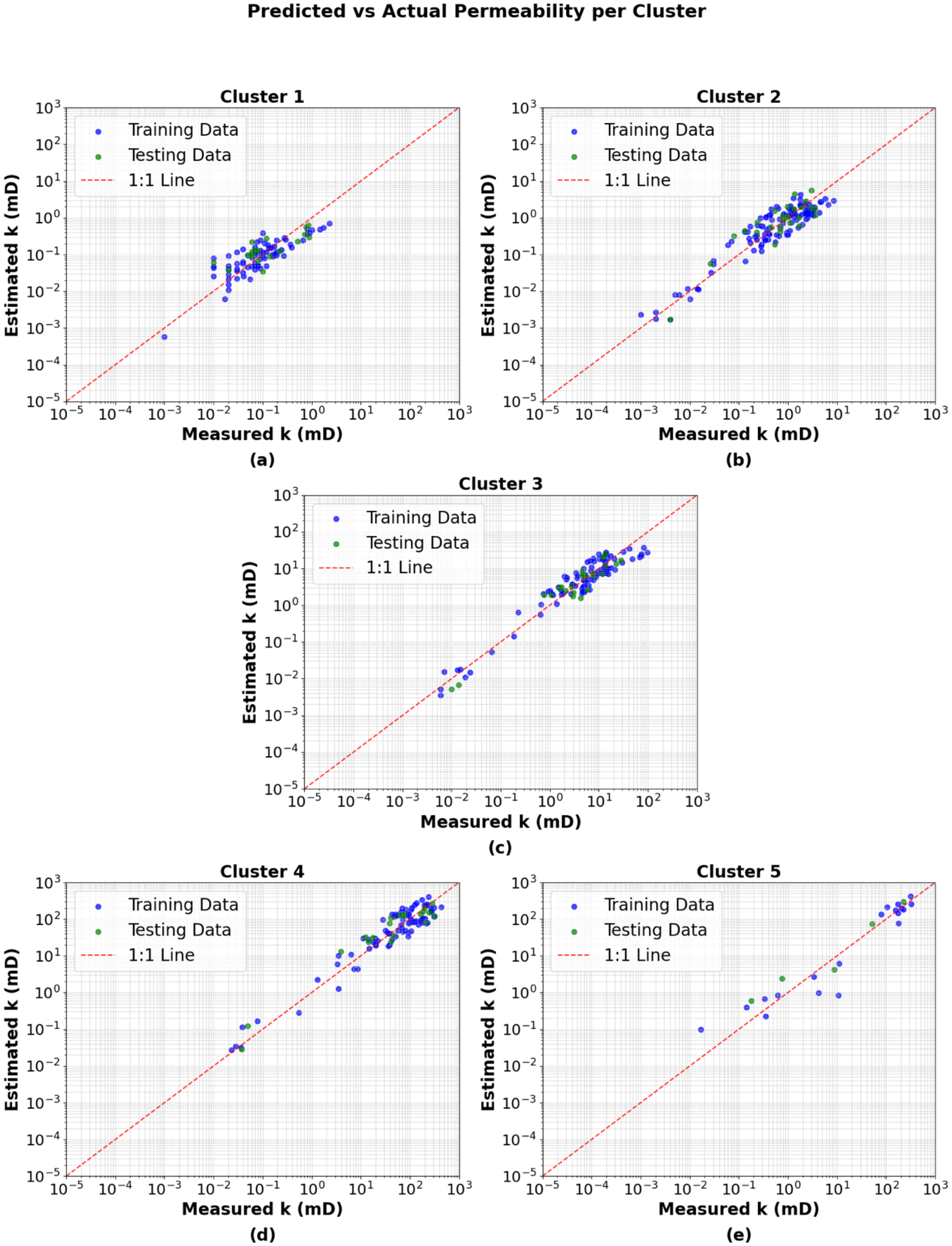

Predicted vs Actual Permeability per Cluster
Cluster 1
Cluster 2
Cluster 3
Cluster 4
Cluster 5
Training Data
Testing Data
1:1 Line
Estimated k (mD)
Measured k (mD)
(a)
(b)
(c)
(d)
(e)

Figure 7. Comparison between measured and estimated permeability values for the training and testing datasets, respectively for each cluster. The dashed red line represents the ideal 1:1 agreement line, where predicted values perfectly match the measured data.

Although the predictive performance varies among clusters (Fig. 7), most clusters exhibit a clear positive relationship between measured and estimated permeability, indicating that the clustering framework captured important pore-structure and flow-related characteristics controlling $k$. Because the FZI rock types were established using measured permeability before the within-cluster data splits, the testing results should be interpreted as conditional evaluations within known rock types rather than fully independent permeability predictions based solely on NMR measurements.

Clusters 1 and 2 correspond primarily to lower-permeability rock types and exhibit relatively narrow ranges of estimated permeability. The small fitted $\beta$ values (Table 1) indicate limited sensitivity of the estimates to $T_{2lm}$ within these clusters. The resulting scatter and restricted estimated-permeability ranges suggest that porosity and a single characteristic relaxation time do not fully represent the variables controlling permeability in these rock types (Fig. 7 vs. Figs. 2d-e). Possible influences include microporosity, disconnected or poorly connected pores, pore-throat restrictions, variations in surface relaxivity, and uncertainty in the fitted parameters. Because these properties were not independently measured, no single pore-scale mechanism can be identified from the optimized exponents alone.

Clusters 3 and 4 show improved agreement between measured and estimated permeability values, with many data points distributed closer to the 1:1 line for both training and testing datasets. These clusters span moderate permeability ranges and demonstrate that the cluster-specific SDR calibration was more successful in capturing the permeability variability within these hydraulic flow units. The broader dynamic range and improved alignment suggest that pore geometry and pore connectivity were more systematically related to the NMR response within these clusters compared to the lower-quality rock types.

Cluster 5 corresponds to the highest-permeability HFU and exhibits the largest permeability ranges. Although the estimated $k$ values generally follow the increasing trend in measured permeability, noticeable scatter remains, particularly for several high-permeability samples. Such deviations are expected in highly heterogeneous carbonate rocks where $k$ may be strongly influenced by vugs, fractures, or preferential flow pathways that are not fully represented by bulk porosity and $T_{2lm}$. Nevertheless, the model for this cluster still preserves the overall permeability trend and provides reasonable first-order estimates across several orders of magnitude.

An important observation from Figure 7 is that the testing datasets generally follow trends similar to those observed for the training datasets, indicating that the cluster-specific calibrations were not solely overfitted to the training data and retained predictive capability for held-out samples within predefined FZI rock types. However, the degree of scatter and cluster-dependent performance also demonstrate that permeability estimation in heterogeneous carbonate reservoirs remains challenging due to substantial variability in pore structure, pore-throat connectivity, surface relaxivity, and characteristic NMR relaxation behavior. Overall, the results confirm that hydraulic rock typing significantly improves the consistency and predictive performance of the modified SDR model relative to applying a single global calibration across the entire heterogeneous dataset.

### 3.4. Effect of rock typing on permeability estimation

Table 2 summarizes the mean RMSLE and CD values calculated across the 10,000 bootstrap calibration iterations for the non-clustered and rock-type-specific modified SDR models. Without rock typing, the model yielded mean RMSLE values of 1.058 and 1.215 for the bootstrap training samples and fixed testing subset, respectively, together with corresponding mean CD values of 0.311 and −0.106. The negative mean testing CD indicates that the globally calibrated model did not adequately represent permeability variability in the fixed testing subset.

Following partitioning into five modeled hydraulic rock types, mean RMSLE ranged from 0.255 to 0.347 for the bootstrap training samples and from 0.240 to 0.427 for the fixed testing subsets. These values were substantially smaller than the corresponding non-clustered values. Mean testing CD ranged from 0.578 to 0.924, compared with -0.106 for the non-clustered model. Because cluster sample sizes and permeability ranges differed, these statistics should not be used to rank the intrinsic predictability of individual rock types directly. Nevertheless, the consistently improved testing statistics indicate that the rock-type-specific calibrations represented the within-cluster permeability relationships more effectively than the global calibration.

Table 2. The calculated root mean square log-transformed error, $RMSLE$, and coefficient of determination, CD, for the training and testing datasets.

| Group | RMSLE (Train) | RMSLE (Test) | CD (Train) | CD (Test) |
|---|---|---|---|---|
| No Clustering | 1.058 | 1.215 | 0.311 | -0.106 |
| Cluster 1 | 0.315 | 0.352 | 0.688 | 0.578 |
| Cluster 2 | 0.293 | 0.327 | 0.867 | 0.754 |
| Cluster 3 | 0.255 | 0.240 | 0.917 | 0.914 |
| Cluster 4 | 0.262 | 0.296 | 0.917 | 0.924 |
| Cluster 5 | 0.347 | 0.427 | 0.919 | 0.851 |
| Cluster 6 | - | - | - | - |

The CD values further demonstrate the improvement associated with hydraulic rock typing. Clusters 3 and 4 exhibited the strongest conditional testing performance, with mean testing CD values of 0.914 and 0.924, respectively. Cluster 5 also showed relatively strong performance, with a mean testing CD of 0.851; however, its fixed testing subset contained only five samples, and this value should therefore be interpreted cautiously.

Clusters 1 and 2 yielded lower but still positive mean testing CD values of 0.578 and 0.754, respectively. The generally similar mean training and testing metrics provide no clear indication of severe overfitting within the fixed data partitions. However, because the same testing subsets were used in every bootstrap iteration, these results quantify sensitivity to bootstrap calibration rather than uncertainty associated with alternative training–testing partitions.

The generally positive CD values for the held-out subsets indicate that the rock-type-specific SDR relationships were transferable to samples withheld from parameter calibration within the same predefined FZI clusters. These results do not, however, evaluate independent assignment of a new sample to a rock type without permeability information.

Figures 8a and 8b compare the measured and estimated permeability depth profiles for the Wellington and Cutter wells using the non-clustered and rock-type-calibrated modified SDR models, respectively. In both wells, the estimated permeability generally follows the major depth intervals associated with measured permeability variations, indicating that the modified SDR framework captures the first-order vertical permeability structure of the reservoir. However, clear differences are observed between the two calibration approaches. We should point out here that the rock-type-specific permeability profiles shown here were generated using FZI rock-type assignments derived from the available core measurements. Therefore, this analysis demonstrates the improvement attainable when rock type is known, rather than a fully log-based field implementation in which rock type is inferred independently.

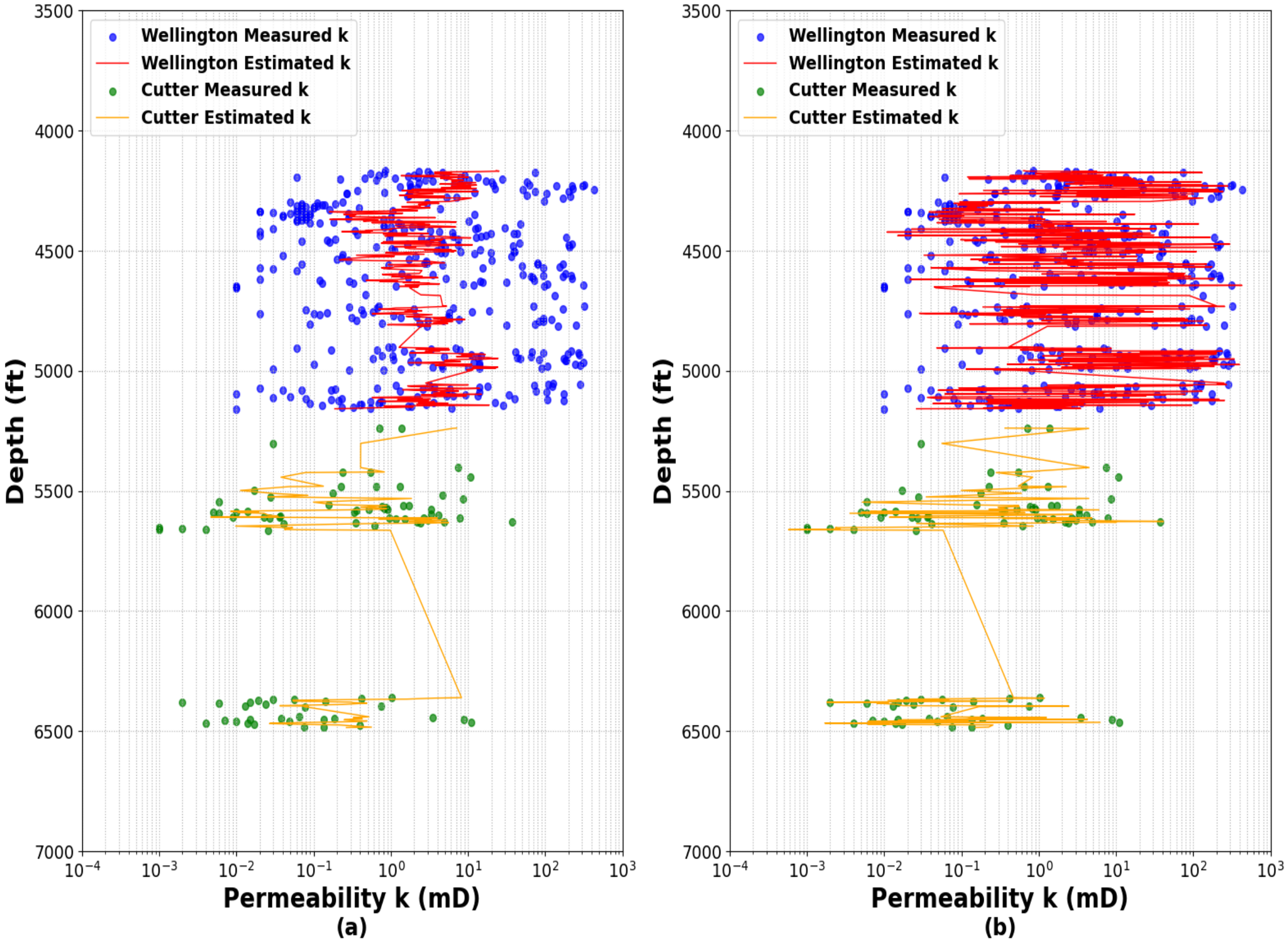

Figure 8. Depth profiles of the measured and estimated permeability via the modified SDR model for two Wellington and Cutter wells from the Arbuckle formation (a) without and (b) with rock typing.

In Figure 8a, corresponding to the non-clustered model, the estimated permeability profiles are relatively compressed and exhibit limited dynamic range, particularly within the Wellington well. The predicted permeability values tend to cluster within narrow intervals and fail to reproduce the full variability observed in the measured data. In contrast, the rock-type-calibrated model shown in Figure 8b produces permeability profiles with substantially broader variability and improved agreement with the measured permeability distribution. The estimated permeability values span a wider range and better reproduce intervals associated with both low- and high-permeability zones. This improvement is particularly evident in the Wellington well, where the estimated permeability profiles extend toward higher permeability values at several depth intervals, more closely matching the measured data distribution. Similar behavior is observed in the Cutter well, especially near depths of approximately 5,500 ft and 6,500 ft, where the clustered model captures stronger permeability contrasts than the non-clustered approach.

The broader spread observed in Figure 8b does not merely reflect increased model variability, but rather indicates that hydraulic rock typing allowed the modified SDR model to adapt to distinct pore systems and flow characteristics within the heterogeneous carbonate reservoir. By calibrating the model separately for different hydraulic flow units, the clustering framework reduced the averaging effects inherent in the global calibration and enabled more realistic representation of permeability heterogeneity across the formation. Overall, the results demonstrate that integrating hydraulic rock typing with NMR log-based permeability estimation significantly improves model performance and provides a more differentiated empirical representation of permeability variability.

A limitation of the present workflow is that FZI-based rock typing requires measured permeability. Consequently, the approach cannot yet be applied as a fully independent NMR permeability-prediction workflow at intervals where permeability data are unavailable. The present results instead quantify the benefit of calibrating the modified SDR model within hydraulically distinct rock types whose identities are known. For field-scale application, an independent rock-typing method based on routinely available measurements—such as NMR relaxation characteristics, conventional well logs, mineralogical information, or image-derived pore-structure descriptors—would be required. Such a method could first assign an unknown interval to a rock type without using permeability, after which the corresponding rock-type-specific SDR parameters could be applied.

### 3.5. Sensitivity of rock typing and SDR calibration to the porosity measurement

To evaluate the sensitivity of the rock-typing and modified SDR calibration results to the source of porosity data, the complete workflow was repeated using core-measured porosity in place of NMR-derived porosity. The core-porosity-based results are presented in the Supporting Information. In both implementations, core-measured permeability was used to calculate FZI; however, the porosity used in the FZI calculation and subsequent modified SDR calibration differed. The main analysis used NMR-derived porosity, whereas the Supporting Information analysis used core-measured porosity.

Substitution of core porosity for NMR porosity changed the calculated RQI, normalized porosity, and FZI values and therefore changed the structure of the resulting HFUs. The NMR-porosity-based analysis identified six clusters, one of which contained only a single sample and was excluded from the SDR calibration, leaving five modeled rock types. In contrast, the core-porosity-based analysis identified eight clusters. This difference is expected because porosity enters both terms used to calculate FZI: $RQI = 0.0314\sqrt{k/\phi}$ and $\phi_z = \phi/(1-\phi)$. Consequently, a change in porosity affects both coordinates of the $RQI$–$\phi_z$ crossplot and can shift samples across FZI-based cluster boundaries. In particular, the generally higher NMR-derived porosity values observed in Fig. 3b tend to produce lower calculated FZI values than the corresponding core-porosity values, although the magnitude of this effect varies among samples.

The two analyses also produced different conditional testing performance. For the five modeled NMR-porosity-based rock types, mean testing RMSLE ranged from 0.240 to 0.427 and mean testing CD ranged from 0.578 to 0.924. For the eight core-porosity-based rock types, mean testing RMSLE ranged from 0.343 to 1.750 and mean testing CD ranged from −1.582 to 0.663 (Table S2). Thus, within the present dataset and calibration procedure, the NMR-porosity-based partition produced more consistently favorable performance across the modeled rock types. The poor performance of some core-porosity-based clusters, particularly Cluster 7, demonstrates that increasing the number of hydraulic flow units does not necessarily produce more stable or transferable SDR calibrations.

Overall, these results demonstrate that FZI-based rock types are not independent of the porosity measurement used to define them. Rock types developed from core porosity should therefore not be transferred directly to an NMR-log-based permeability workflow without evaluating core–log porosity consistency and, where necessary, establishing an appropriate reconciliation or calibration procedure. The present results also emphasize that field application of the proposed framework requires representative core data and careful matching of the porosity measurement used for rock-type definition with that used in the subsequent SDR calibration.

## 4. Conclusion

In this study, we demonstrated that FZI-based rock typing improves the calibration and conditional testing performance of the modified SDR model within predefined hydraulic flow units in the heterogeneous carbonate formations of the Arbuckle formation, Kansas. Using core measurements and NMR well log data, 470 matched samples were analyzed and classified into six distinct hydraulic flow units through FZI-based K-means clustering. The clustering results revealed pronounced variability in pore structure, pore-throat connectivity, and permeability behavior across the reservoir, highlighting the strong heterogeneity characteristic of carbonate systems. Results showed that a single global SDR calibration is insufficient to adequately represent the complex petrophysical behavior of the Arbuckle carbonate formation. In contrast, calibrating the modified SDR model separately for each hydraulic rock type significantly improved permeability estimation for both the training and testing datasets. The rock-type-specific models produced lower RMSLE values and generally higher CD values compared to the non-clustered approach, indicating improved agreement between measured and estimated permeability values. The cluster-specific calibrations also better reproduced the permeability variability observed in the Wellington and Cutter wells, demonstrating the importance of accounting for hydraulic heterogeneity when applying NMR-based permeability models in carbonate reservoirs. The optimized modified SDR parameters varied substantially among the hydraulic flow units, demonstrating that they are not universal across the Arbuckle carbonate formation. Several rock types yielded $T_{2lm}$ exponents close to zero, indicating that $T_{2lm}$ contributed little additional information to permeability estimation within those groups. Rather than assigning these exponents a unique pore-scale interpretation, this result should be viewed as evidence that the SDR functional form has limited applicability to some heterogeneous carbonate rock types. The fitted parameters may reflect combined effects of pore-system complexity, surface-relaxivity variability, correlations among model predictors, restricted within-cluster variability, and calibration uncertainty. Thus, this study demonstrates that hydraulic rock typing can improve the empirical calibration of the modified SDR relationship relative to a single formation-wide calibration; it does not establish that the SDR model fully represents the physical controls on permeability in heterogeneous carbonates. The resulting parameters should therefore be treated as formation- and rock-type-specific calibration quantities rather than universal physical properties. Future studies should incorporate independent measurements of pore structure, surface relaxivity, and formation factor and compare the rock-type-specific SDR approach with alternative permeability models using larger multiwell datasets. Because FZI was calculated using measured permeability, the reported testing results represent performance for samples with known FZI rock-type membership. Future work should also develop and validate a permeability-independent rock-typing procedure before the proposed framework can be used for blind permeability prediction from NMR measurements.

**Acknowledgement**

BG is grateful to the University of Texas at Arlington for financial support through faculty startup fund and the STARs award. The authors used AI-assisted language editing tools (ChatGPT 5.6) to improve grammar and readability. All scientific interpretations and conclusions remain the responsibility of the authors.

**Authors' Note**

The authors do not have any conflicts of interest or financial disclosures to report.

**Data Availability**

The dataset analyzed here is available at (Lokesh et al., 2026):

http://www.hydroshare.org/resource/308497f85d3344d4855c275ca71aa0de

The code used in this study is available at:

https://github.com/B1a-z3/Combining-Rock-Typing-and-NMR-Logging-to-Estimate-Permeability-in-Complex-Carbonate-Formations

# Supporting Information

The Supporting Information analysis differs from the main analysis in the porosity used for FZI-based rock typing: core-measured porosity is used here, whereas NMR-derived porosity is used in the main manuscript. In our analyses, rock typing precedes the within-cluster 80/20 splits and requires core-measured permeability. Fig. S1a shows the results of the elbow method and K = 8 (eight clusters or rock types), the selected number of hydraulic flow units (HFUs). The resulting clusters are shown on the log-log plot of reservoir quality index (RQI) versus normalized porosity ($\phi_z$) in Fig. S1b. We found FZI = 0.451, 1.034, 2.271, 4.720, 11.196, 28.960, 78.957, and 282.887 respectively for clusters 1, 2, 3, 4, 5, 6, 7, and 8. Figure S2 presents the relationship between permeability and porosity for the identified HFUs. The distributions of the optimized modified SDR model parameters for the eight hydraulic rock types are presented in Figure S3. The corresponding average values of $C_{mSDR}$, $\alpha$, $\beta$, and FZI are summarized in Table S1. Figure S4 presents the comparison between laboratory-measured permeability and permeability estimated using the cluster-specific modified SDR models for both the training and testing datasets (a representative bootstrap realization selected as the iteration whose training RMSLE was closest to the mean training RMSLE across the 10,000 iteration). Table S2 summarizes the RMSLE and CD values obtained for the training and testing datasets using both the non-clustered approach and the rock-type-specific modified SDR models. Figures S5a and S5b compare the measured and estimated permeability depth profiles for the Wellington and Cutter wells using the non-clustered and rock-type-calibrated modified SDR models, respectively.

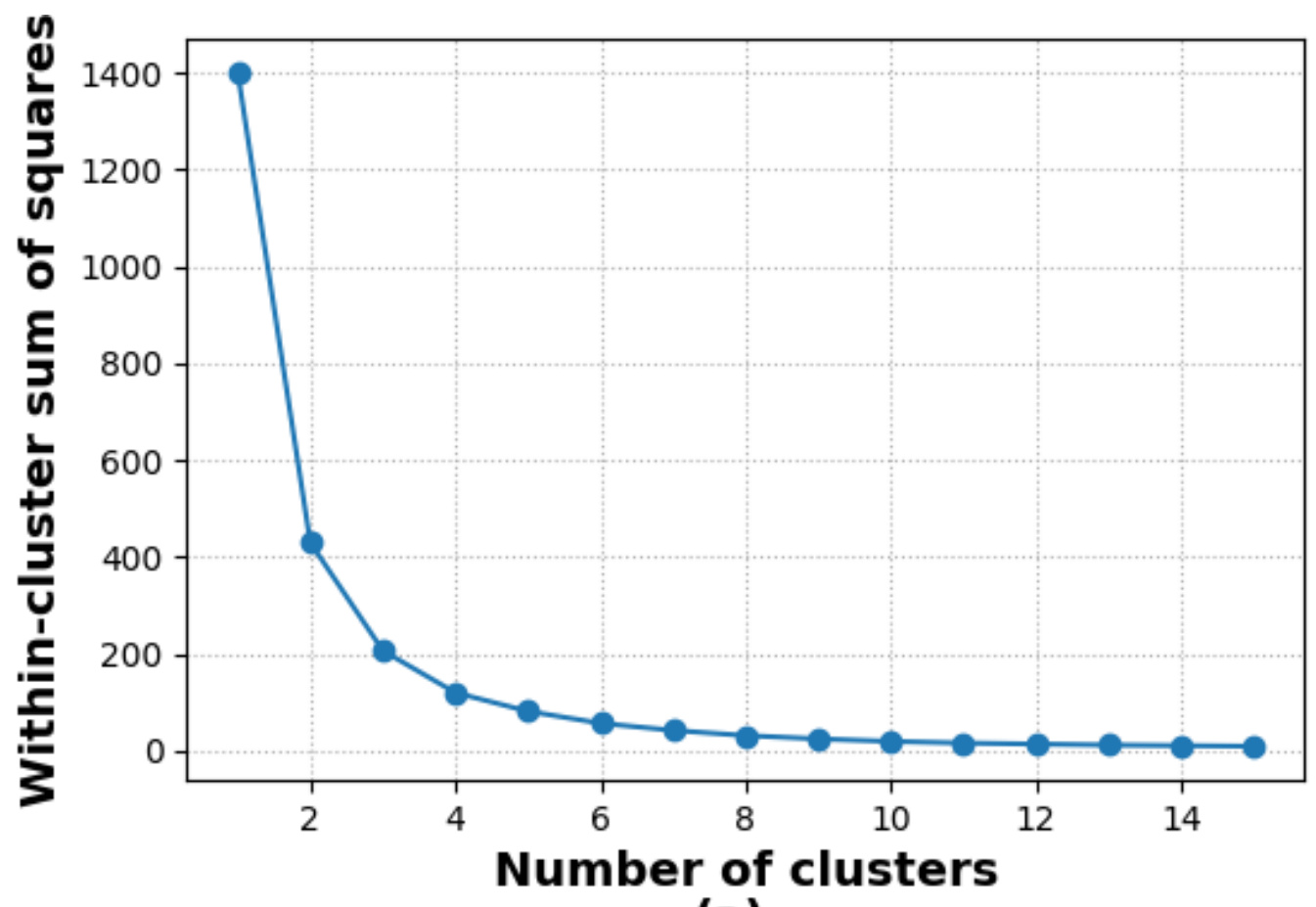

(a)

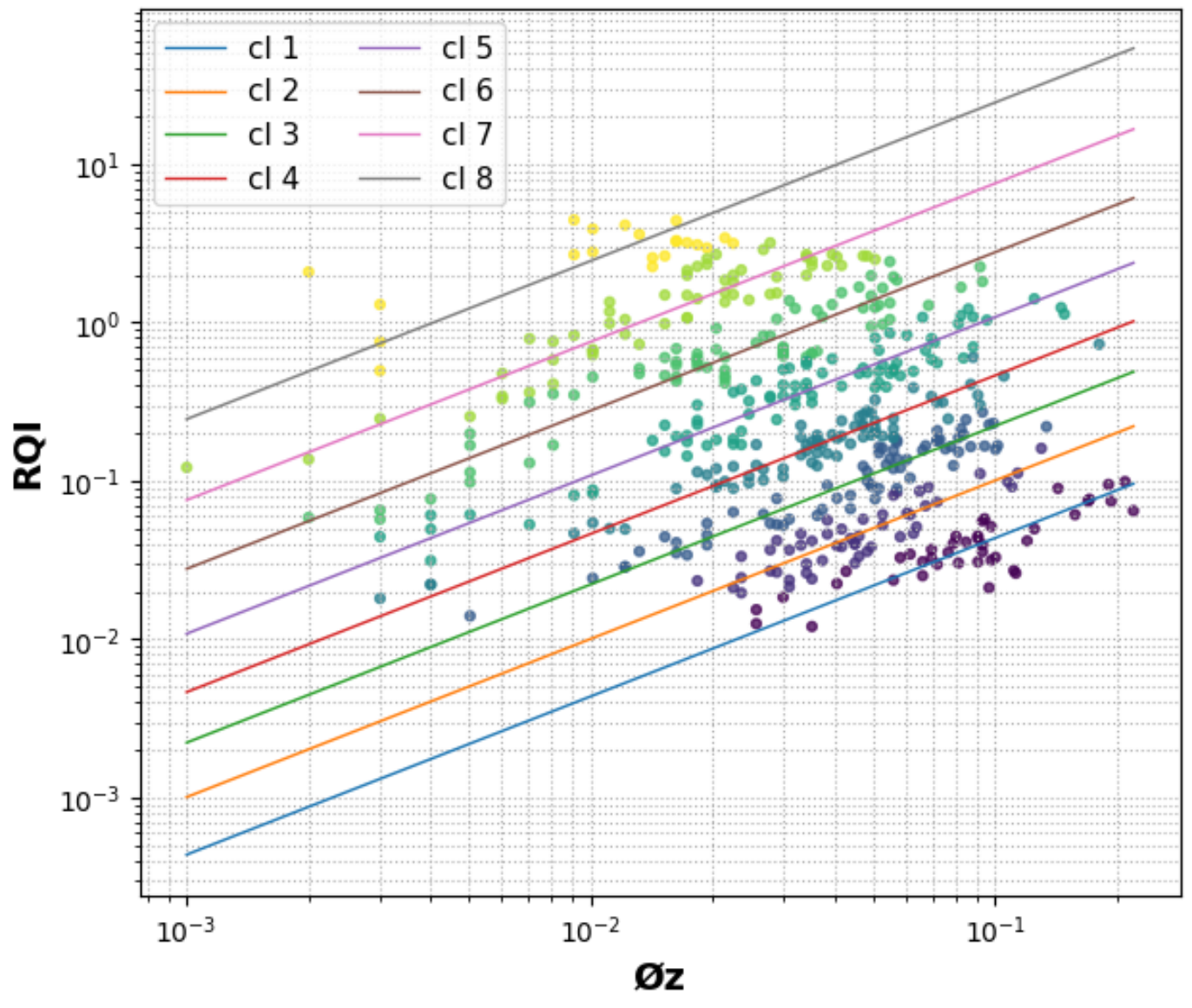

(b)

Figure S1. (a) Elbow plot of within-cluster sum of squares versus number of clusters used to determine the optimal K = 8 for K-means clustering on log(FZI), and (b) the resulting eight

hydraulic flow unit clusters displayed on a log-log RQI versus $\phi_z$ cross plot with corresponding regression lines.

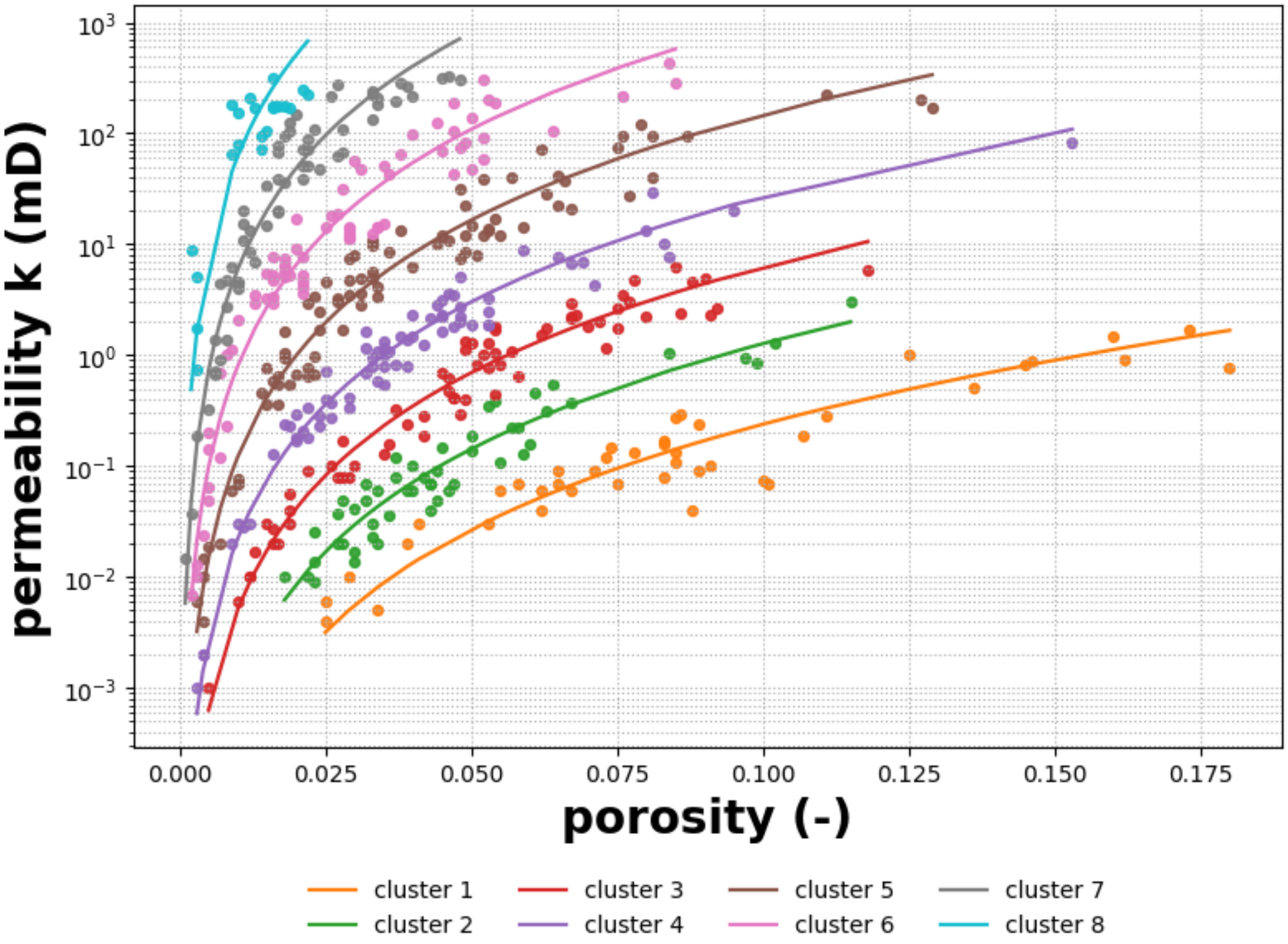

Figure S2. Permeability versus core porosity and the corresponding fit of the modified Kozeny-Carman model, Eq. (4), for eight hydraulic flow unit clusters illustrating the distinct permeability-porosity relationships captured within each cluster.

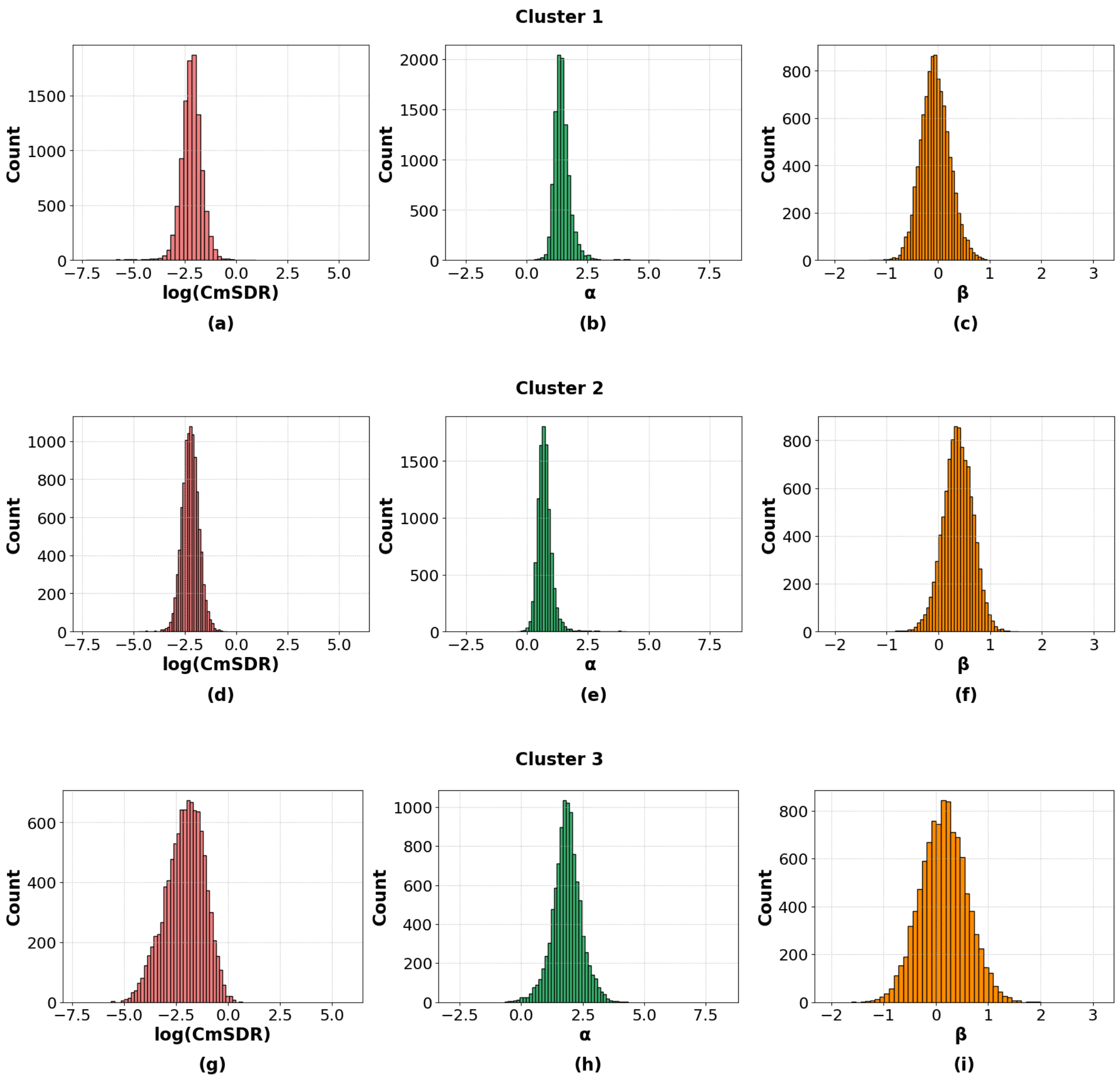

Figure S3. Distribution of the optimized SDR model parameters, including $\log(C_{mSDR})$, $\alpha$, and $\beta$, showing the frequency and stability of the parameter values obtained during model calibration for each cluster and rock type.

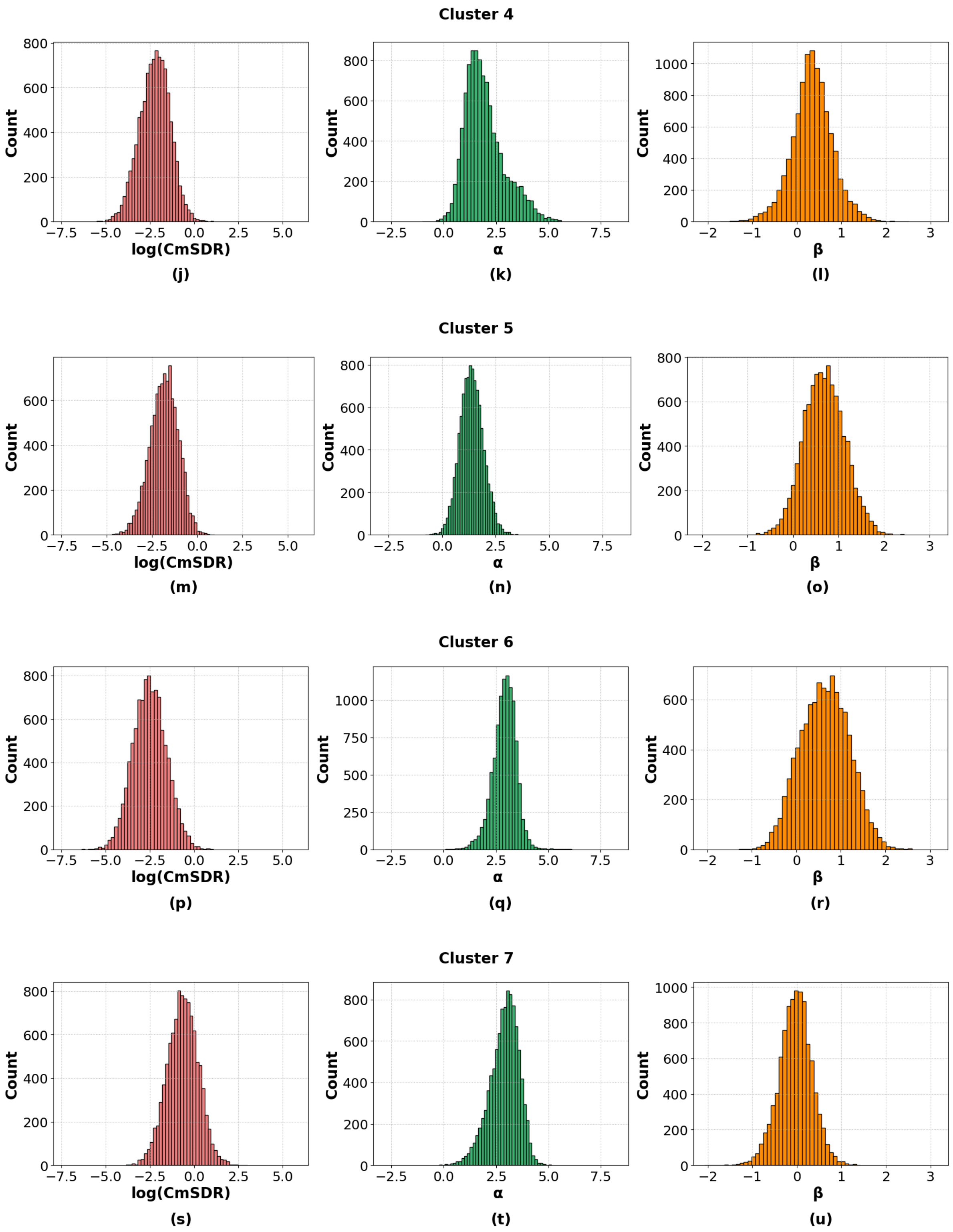

Figure S3. Continued.

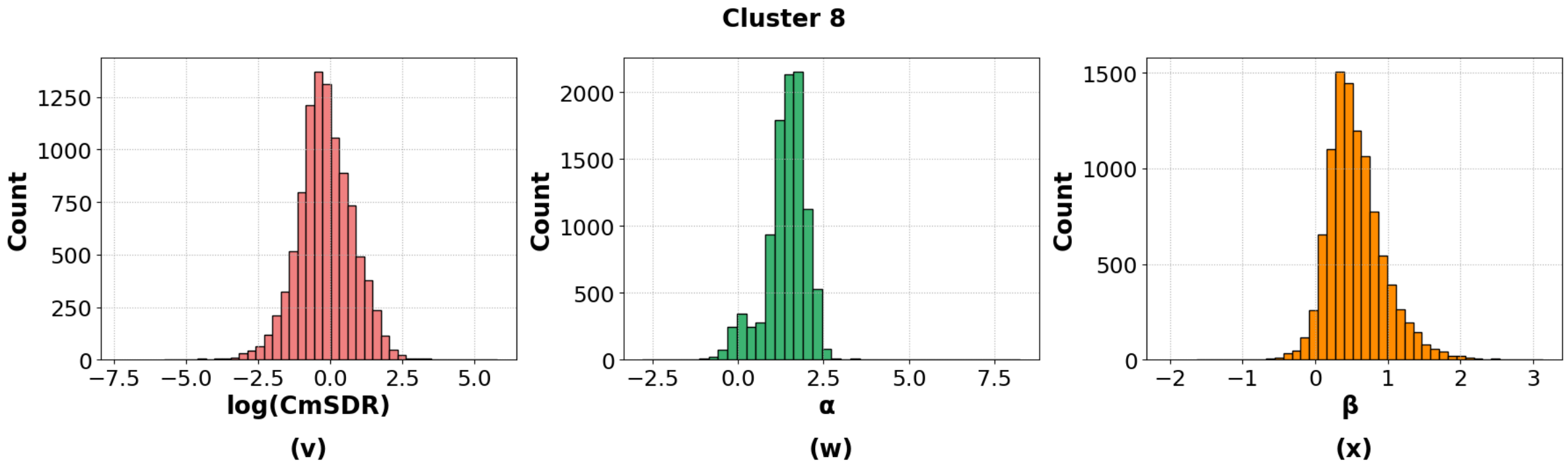

Figure S3. Continued.

Table S1. The modified SDR model parameters optimized using carbonate rock samples from the Arbuckle formation, Kansas.

| **Group** | **No. of samples** | $C_{mSDR}$ **(mD ms$^{-\beta}$)** | $\alpha$ | $\beta$ | **FZI** |
|---|---|---|---|---|---|
| No Clustering | 470 | 0.0006 ± 0.000 | 0.924 ± 0.191 | 1.469 ± 0.156 | N/A |
| Cluster 1 | 42 | 0.0153 ± 0.323 | 1.452 ± 0.353 | -0.051 ± 0.276 | 0.451 ± 0.110 |
| Cluster 2 | 51 | 0.0090 ± 0.005 | 0.721 ± 0.317 | 0.374 ± 0.285 | 1.034 ± 0.237 |
| Cluster 3 | 68 | 0.0466 ± 0.398 | 1.847 ± 0.579 | 0.148 ± 0.443 | 2.271 ± 0.454 |
| Cluster 4 | 73 | 0.0371 ± 0.094 | 1.949 ± 0.951 | 0.343 ± 0.454 | 4.720 ± 0.887 |
| Cluster 5 | 83 | 0.0601 ± 0.323 | 1.364 ± 0.547 | 0.684 ± 0.448 | 11.196 ± 2.878 |
| Cluster 6 | 73 | 0.0302 ± 0.492 | 2.912 ± 0.553 | 0.636 ± 0.541 | 28.960 ± 8.248 |
| Cluster 7 | 59 | 1.6853 ± 7.855 | 2.884 ± 0.695 | -0.035 ± 0.370 | 78.957 ± 23.933 |
| Cluster 8 | 21 | 82.512 ± 44.690 | 1.400 ± 0.615 | 0.556 ± 0.389 | 282.887 ± 199.529 |

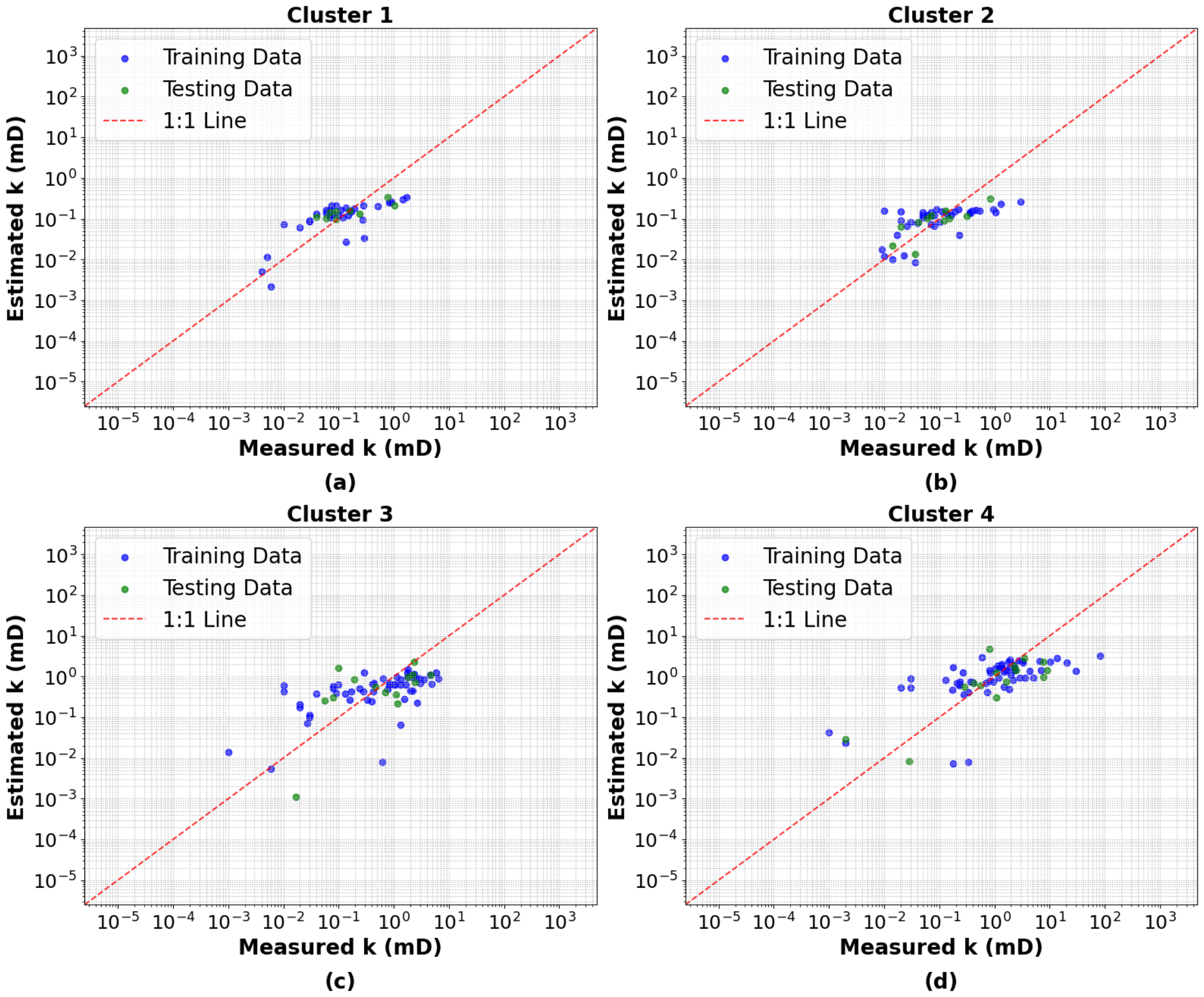

Figure S4. Comparison between measured and estimated permeability values for the training and testing datasets, respectively for each cluster. The dashed red line represents the ideal 1:1 agreement line, where predicted values perfectly match the measured data.

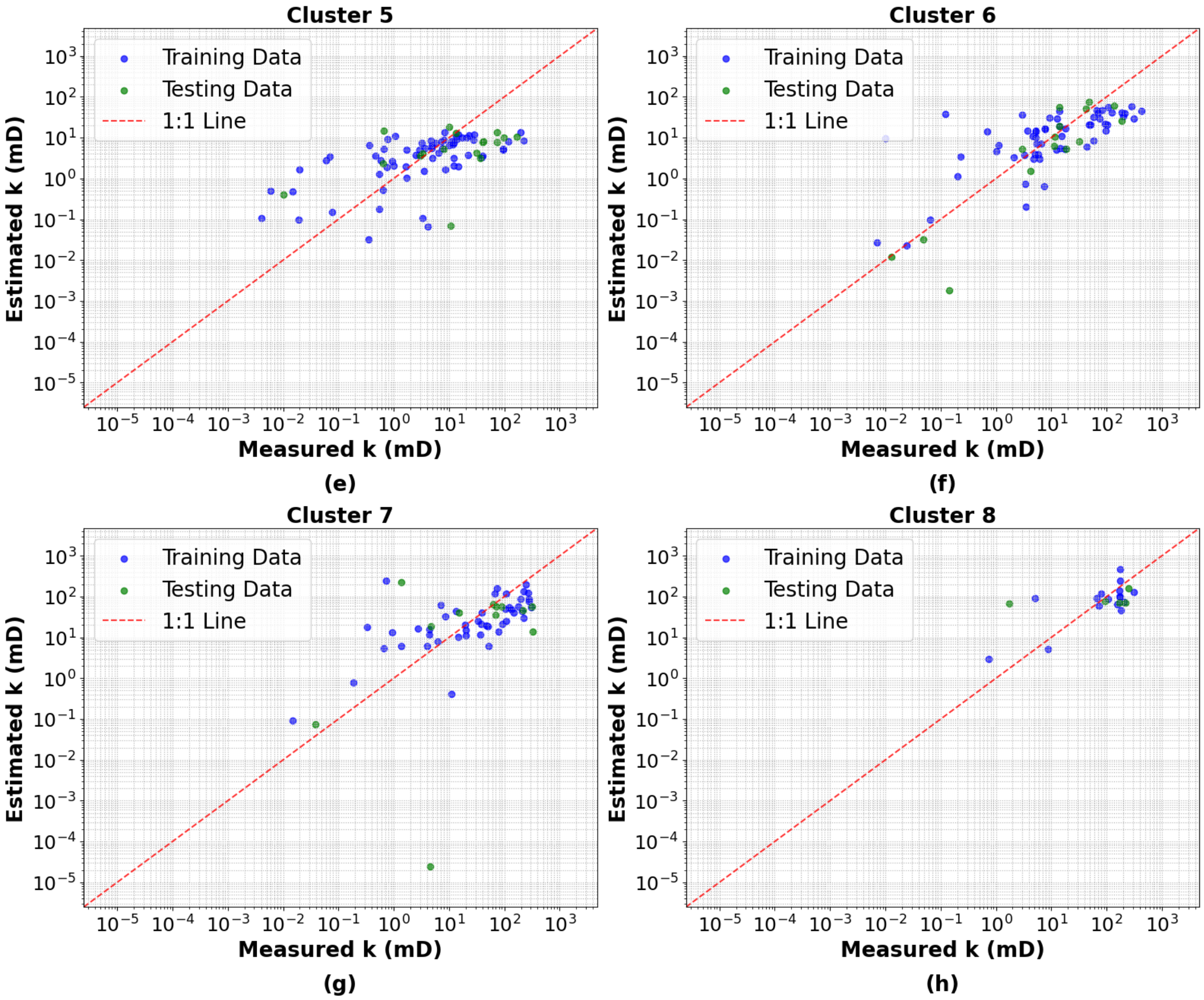

Figure S4. Continued.

Table S2. Mean RMSLE and CD values calculated across 10,000 bootstrap calibration iterations.

| Group | RMSLE (Train) | RMSLE (Test) | CD (Train) | CD (Test) |
|---|---|---|---|---|
| No Clustering | 1.058 | 1.215 | 0.311 | -0.106 |
| Cluster 1 | 0.423 | 0.361 | 0.552 | 0.386 |
| Cluster 2 | 0.463 | 0.343 | 0.397 | 0.490 |
| Cluster 3 | 0.669 | 0.697 | 0.372 | 0.027 |
| Cluster 4 | 0.637 | 0.641 | 0.419 | 0.517 |
| Cluster 5 | 0.803 | 1.005 | 0.377 | 0.023 |
| Cluster 6 | 0.743 | 0.677 | 0.505 | 0.663 |
| Cluster 7 | 0.686 | 1.750 | 0.456 | -1.582 |
| Cluster 8 | 0.391 | 0.793 | 0.562 | 0.047 |

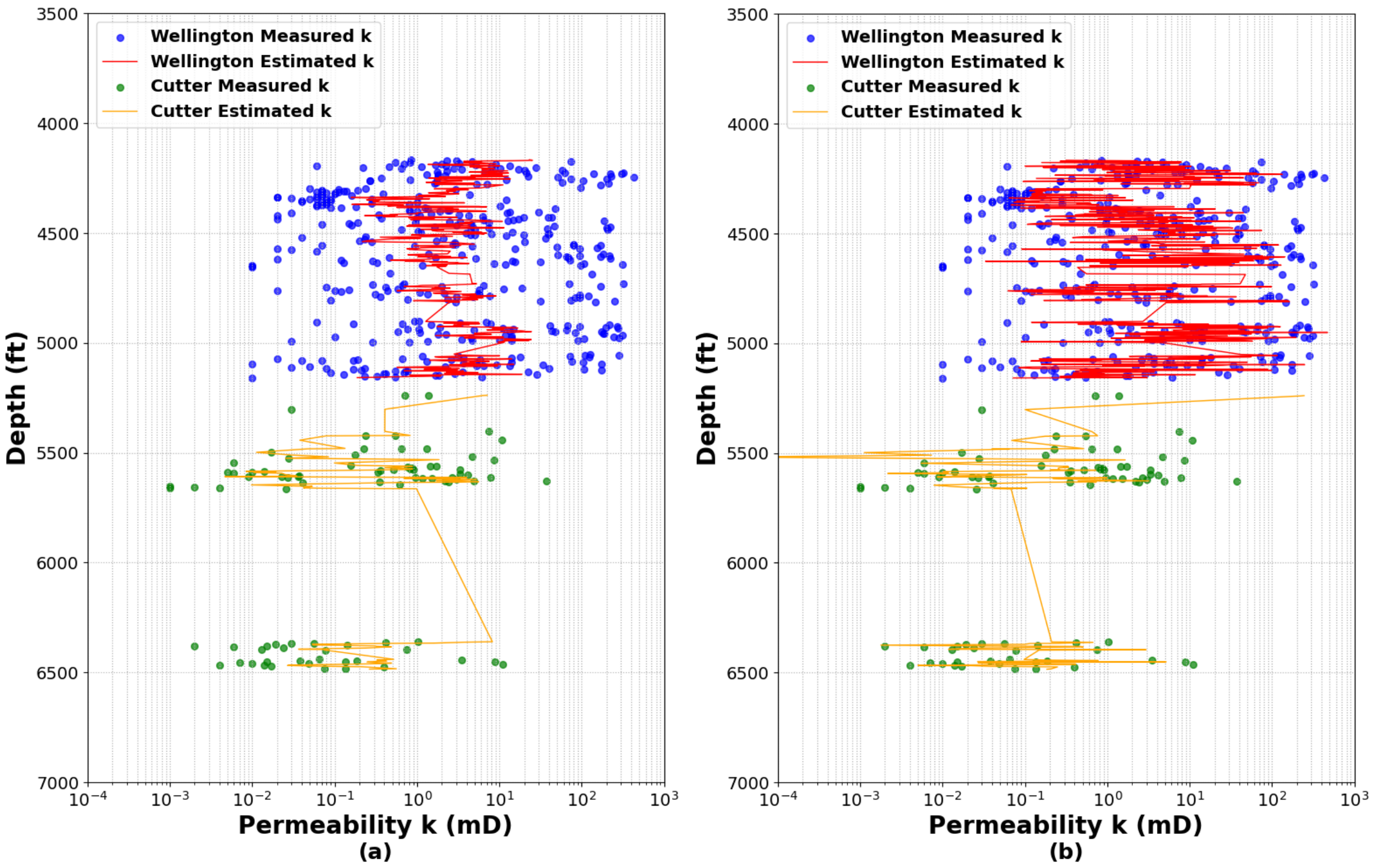

Figure S5. Depth profiles of the measured and estimated permeability via the modified SDR model for two Wellington and Cutter wells from the Arbuckle formation (a) without and (b) with rock typing.